%% this file is bogard9.tex
 
%\input fiat
%%%%%%%%%%%%%%%%%%%%%%%%%%%%%%%%%%%%%%%%%% FORMATO
\newcount\mgnf\newcount\tipi\newcount\tipoformule
\newcount\aux\newcount\driver
\mgnf=0          %ingrandimento
\tipoformule=0   %=0 da numeroparagrafo.numeroformula; =1 numero
                 %scritto
\aux=0           %=0 non produce .aux per le formule, =1 si. 1 da anche
                 %la data a pie' pagina
           \def\9#1{\ifnum\aux=1#1\else\relax\fi}
%%%% INCIPIT
\ifnum\mgnf=0
   \magnification=\magstep0
%\hsize=11.5truecm\vsize=19.5truecm%stampa
\hsize=14.5truecm\vsize=21truecm%preprint
   \parindent=4.pt\fi
\ifnum\mgnf=1
   \magnification=\magstep1
\hsize=16.0truecm\vsize=22.5truecm\baselineskip14pt\vglue6.3truecm%mpej
%\hsize=11.5truecm\vsize=19.5truecm\baselineskip14pt%stampa
%\hsize=15.5truecm\vsize=22.truecm\baselineskip=15pt%preprint
   \parindent=4.pt\fi
%%%%%%%%%%%
%\overfullrule=10pt
%
%%%%%GRECO%
\let\a=\alpha   \let\g=\gamma    \let\d=\delta \let\e=\varepsilon
\let\z=\zeta     \let\th=\vartheta\let\k=\kappa \let\l=\lambda
\let\m=\mu    \let\n=\nu    \let\x=\xi       \let\p=\pi    \let\r=\rho
\let\s=\sigma \let\t=\tau    \let\f=\varphi
   
 \let\D=\Delta    \let\L=\Lambda

%%%%%%%%%%%%%%%%%%%%%  NUMERAZIONE PAGINE
{\count255=\time\divide\count255 by 60 \xdef\oramin{\number\count255}
        \multiply\count255 by-60\advance\count255 by\time
   \xdef\oramin{\oramin:\ifnum\count255<10 0\fi\the\count255}}
\def\ora{\oramin }
 
\def\data{\number\day/\ifcase\month\or gennaio \or febbraio \or marzo \or
aprile \or maggio \or giugno \or luglio \or agosto \or settembre
\or ottobre \or novembre \or dicembre \fi/\number\year;\ \ora}
 
\setbox200\hbox{$\scriptscriptstyle \data $}
 
\newcount\pgn \pgn=1
\def\foglio{\number\numsec:\number\pgn
\global\advance\pgn by 1}
\def\foglioa{A\number\numsec:\number\pgn
\global\advance\pgn by 1}
 
%%%%%%%%%%%%%%%%% EQUAZIONI CON NOMI SIMBOLICI
%%%
%%% per assegnare un nome simbolico ad una equazione basta
%%% scrivere \Eq(...) o, in \eqalignno, \eq(...) o,
%%% nelle appendici, \Eqa(...) o \eqa(...):
%%% dentro le parentesi e al posto dei ...
%%% si puo' scrivere qualsiasi commento;
%%% per assegnare un nome simbolico ad una figura, basta scrivere
%%% \geq(...); per avere i nomi
%%% simbolici segnati a sinistra delle formule e delle figure si deve
%%% dichiarare il documento come bozza, iniziando il testo con
%%% \BOZZA.
%%% All' inizio di ogni paragrafo si devono definire il
%%% numero del paragrafo e della prima formula dichiarando
%%% \numsec=... \numfor=...  (brevetto Eckmannn); all'inizio del lavoro
%%% bisogna porre \numfig=1 (il numero delle figure non contiene la sezione.
%%% Si possono citare formule o figure seguenti; le corrispondenze fra nomi
%%% simbolici e numeri effettivi sono memorizzate nel file \jobname.aux, che
%%% viene letto all'inizio, se gia' presente. E' possibile citare anche
%%% formule o figure che appaiono in altri file, purche' sia presente il
%%% corrispondente file .aux.
%%%%%%%%%%%%%%%%%%%%%%%%%%%%%%%%%%%
 
\global\newcount\numsec\global\newcount\numfor
\global\newcount\numfig
\gdef\profonditastruttura{\dp\strutbox}
\def\senondefinito#1{\expandafter\ifx\csname#1\endcsname\relax}
\def\SIA #1,#2,#3 {\senondefinito{#1#2}
\expandafter\xdef\csname #1#2\endcsname{#3} \else
\write16{???? ma #1,#2 e' gia' stato definito !!!!} \fi}
\def\etichetta(#1){(\veroparagrafo.\veraformula)
\SIA e,#1,(\veroparagrafo.\veraformula)
 \global\advance\numfor by 1
\9{\write15{\string\FU (#1){\equ(#1)}}}
\9{ \write16{ EQ \equ(#1) == #1  }}}
\def \FU(#1)#2{\SIA fu,#1,#2 }
\def\etichettaa(#1){(A\veroparagrafo.\veraformula)
 \SIA e,#1,(A\veroparagrafo.\veraformula)
 \global\advance\numfor by 1
\9{\write15{\string\FU (#1){\equ(#1)}}}
\9{ \write16{ EQ \equ(#1) == #1  }}}
\def\getichetta(#1){Fig. \verafigura
 \SIA e,#1,{\verafigura}
 \global\advance\numfig by 1
\9{\write15{\string\FU (#1){\equ(#1)}}}
\9{ \write16{ Fig. \equ(#1) ha simbolo  #1  }}}
\newdimen\gwidth
\def\BOZZA{
\def\alato(##1){
 {\vtop to \profonditastruttura{\baselineskip
 \profonditastruttura\vss
 \rlap{\kern-\hsize\kern-1.2truecm{$\scriptstyle##1$}}}}}
\def\galato(##1){ \gwidth=\hsize \divide\gwidth by 2
 {\vtop to \profonditastruttura{\baselineskip
 \profonditastruttura\vss
 \rlap{\kern-\gwidth\kern-1.2truecm{$\scriptstyle##1$}}}}}
\footline={\rlap{\hbox{\copy200}\ $\st[\number\pageno]$}\hss\tenrm
\folio\hss}}
%\foglio\hss}}  per avere il numero paragrafo.paginadi paragrafo
\def\alato(#1){}
\def\galato(#1){}
\def\veroparagrafo{\number\numsec}\def\veraformula{\number\numfor}
\def\verafigura{\number\numfig}
\def\geq(#1){\getichetta(#1)\galato(#1)}
\def\Eq(#1){\eqno{\etichetta(#1)\alato(#1)}}
\def\eq(#1){\etichetta(#1)\alato(#1)}
\def\Eqa(#1){\eqno{\etichettaa(#1)\alato(#1)}}
\def\eqa(#1){\etichettaa(#1)\alato(#1)}
\def\eqv(#1){\senondefinito{fu#1}$\clubsuit$#1\write16{Manca #1 !}%
\else\csname fu#1\endcsname\fi}
\def\equ(#1){\senondefinito{e#1}\eqv(#1)\else\csname e#1\endcsname\fi}
 
\openin13=#1.aux \ifeof13 \relax \else
\input #1.aux \closein13\fi
\openin14=\jobname.aux \ifeof14 \relax \else
\input \jobname.aux \closein14 \fi
\9{\openout15=\jobname.aux}
%%%%%%%%%%%%%%%%% CARATTERI
\newskip\ttglue
\font\dodiciit=cmti12
\font\titolo=cmbx12 scaled \magstep2
%\font\dodicirm=cmr10 scaled \magstep1
%\font\dodicibf=cmbx10 scaled \magstep1
%\font\dodiciit=cmti10 scaled \magstep1
%  \font\titolo=cmbx10 scaled \magstep1
  \font\eighttt=cmtt8 \font\sevenit=cmti7  \font\sevensl=cmsl8
%%%%%%%
\def\settepunti{\def\rm{\fam0\sevenrm}
\textfont0=\sevenrm \scriptfont0=\fiverm \scriptscriptfont0=\fiverm
\textfont1=\seveni \scriptfont1=\fivei   \scriptscriptfont1=\fivei
\textfont2=\sevensy \scriptfont2=\fivesy   \scriptscriptfont2=\fivesy
\textfont3=\tenex \scriptfont3=\tenex   \scriptscriptfont3=\tenex
\textfont\itfam=\sevenit  \def\it{\fam\itfam\sevenit}%
\textfont\slfam=\sevensl  \def\sl{\fam\slfam\sevensl}%
\textfont\ttfam=\eighttt  \def\tt{\fam\ttfam\eighttt}
\textfont\bffam=\sevenbf  \scriptfont\bffam=\fivebf
\scriptscriptfont\bffam=\fivebf  \def\bf{\fam\bffam\sevenbf}%
\tt \ttglue=.5em plus.25em minus.15em
\setbox\strutbox=\hbox{\vrule height6.5pt depth1.5pt width0pt}%
\normalbaselineskip=8pt\let\sc=\fiverm \normalbaselines\rm}
%\catcode`@=11      %%% ridefinizione di \footnote per note piccole
%\def\footnote#1{\edef\@sf{\spacefactor\the\spacefactor}#1\@sf
%\insert\footins\bgroup\settepunti
% \interlinepenalty100 \let\par=\endgraf
%   \leftskip=0pt \rightskip=0pt
%   \splittopskip=9pt plus 1pt minus 1pt \floatingpenalty=20000
%   \smallskip\item{#1}\bgroup\strut\aftergroup\@foot\let\next}
%\skip\footins=11pt plus 1.8pt minus 3.5pt
%\dimen\footins=29pc
%\catcode`@=12
%
\let\nota=\settepunti

\newdimen\xshift \newdimen\xwidth \newdimen\yshift  \newdimen\ywidth
 
\def\ins#1#2#3{\vbox to0pt{\kern-#2 \hbox{\kern#1 #3}\vss}\nointerlineskip}
 
\def\eqfig#1#2#3#4#5{
\par\xwidth=#1 \xshift=\hsize \advance\xshift
by-\xwidth \divide\xshift by 2
\yshift=#2 \divide\yshift by 2
\line{\hglue\xshift \vbox to #2{\vfil
#3 \includegraphics{#4.ps}
}\hfill\raise\yshift\hbox{#5}}}
 
\def\eqfigbis#1#2#3#4#5#6#7{
\par\xwidth=#1 \multiply\xwidth by 2
\xshift=\hsize \advance\xshift
by-\xwidth \divide\xshift by 3
\yshift=#2 \divide\yshift by 2
\ywidth=#2
\line{\hglue\xshift
\vbox to \ywidth{\vfil #3 \includegraphics{#4.ps}}
\hglue30pt
\vbox to \ywidth{\vfil #5 \includegraphics{#6.ps}}
\hfill\raise\yshift\hbox{#7}}}

\def\dimenfor#1#2{\par\xwidth=#1 \multiply\xwidth by 2
\xshift=\hsize \advance\xshift
by-\xwidth \divide\xshift by 3
\divide\xwidth by 2
\yshift=#2 %\divide\yshift by 2
\ywidth=#2}

\def\eqfigfor#1#2#3#4#5#6#7#8#9{
\line{\hglue\xshift
\vbox to \ywidth{\vfil #1 \includegraphics{#2.ps}}
\hglue30pt
\vbox to \ywidth{\vfil #3 \includegraphics{#4.ps}}\hfill}
%\vglue20pt
\line{\hfill\hbox{#9}}
\line{\hglue\xshift
\vbox to \ywidth{\vfil #5 \includegraphics{#6.ps}}
\hglue30pt
\vbox to\ywidth {\vfil #7 \includegraphics{#8.ps}}\hfill}}
%\raise\yshift\hbox{#9}}

\def\eqfigter#1#2#3#4#5#6#7{
\line{\hglue\xshift
\vbox to \ywidth{\vfil #1 \includegraphics{#2.ps}}
\hglue30pt
\vbox to \ywidth{\vfil #3 \includegraphics{#4.ps}}\hfill}
\multiply\xshift by 3 \advance\xshift by \xwidth \divide\xshift by 2
\line{\hfill\hbox{#7}}
\line{\hglue\xshift
\vbox to \ywidth{\vfil #5 \includegraphics{#6.ps}}\hfill}}

%%%%% DISEGNO
% permette di scrivere file postscript in un test tex:
% il file postcript va scritto nel file tex riga per riga nella forma
% \8< .......... >
% dove i puntini stanno per una riga del file. A queste righe va
% premesso \figini#1 e posposto \figfin
% Il file postscript riceve il  nome #1.ps
%%%%%
 
\def\8{\write13}

%%%%%% DEFINIZIONI VARIE
\def\didascalia#1{\vbox{\nota\0#1\hfill}\vskip0.3truecm}
 
\def\V#1{{\,\underline#1\,}}
\def\T#1{#1\kern-4pt\lower9pt\hbox{$\widetilde{}$}\kern4pt{}}
\let\dpr=\partial\def\Dpr{{\V\dpr}}
\let\io=\infty\let\ig=\int
\def\fra#1#2{{#1\over#2}}\def\media#1{\langle{#1}\rangle}\let\0=\noindent
 
\def\guida{\leaders\hbox to 1em{\hss.\hss}\hfill}
\def\tende#1{\vtop{\ialign{##\crcr\rightarrowfill\crcr
              \noalign{\kern-1pt\nointerlineskip}
              \hglue3.pt${\scriptstyle #1}$\hglue3.pt\crcr}}}
\def\otto{\,{\kern-1.truept\leftarrow\kern-5.truept\to\kern-1.truept}\,}

\def\pagina{\vfill\eject}
\let\ciao=\bye
\def\txt{\textstyle}
\def\st{\scriptscriptstyle}
\def\*{\vskip0.3truecm}
\def\lis#1{{\overline #1}}

%%%%%%%LATINORUM
\def\etc{\hbox{\it etc}}
\def\ap{\hbox{\it a priori\ }}\def\aps{\hbox{\it a posteriori\ }}
\def\ie{\hbox{\it i.e.\ }}
\def\fiat{{}}
%%%%%%%DEFINIZIONI LOCALI
\def\annota#1{\footnote{${}^#1$}}
%{{\parindent=0pt%\nota\0#2\vfill}}}
 
\ifnum\aux=1\BOZZA\else\relax\fi
\ifnum\tipoformule=1\let\Eq=\eqno\def\eq{}\let\Eqa=\eqno\def\eqa{}
\def\equ{{}}\footline={\hss\tenrm\folio\hss}\fi
 
\def\CC{{\cal C}}\def\FF{{\cal F}}
\def\DEF{\,{\buildrel def \over =}\,}\let\==\equiv
\def\2{\kern8.mm\hfill}\def\3{\cr\noalign{\vskip1.mm}}
\def\?{\hbox{$\clubsuit$}}\def\\{\hfill\break}\def\bottone{\hbox{$\bullet$}}
\def\paginazero{\ifnum\mgnf=0\vfill\eject\else\relax\fi}
\def\paginauno{\ifnum\mgnf=1\vfill\eject\else\relax\fi}

\fiat
%\headline{\nota\hss DRAFT 9: NOT FOR CIRCULATION}
%\vglue6cm
\centerline{\titolo Chaotic principle: an experimental test}
\vskip1.truecm
\def\1{\ifnum\mgnf=0\pagina\fi}
 
\centerline{\dodiciit F. Bonetto{$^*$}, G.
Gallavotti{$^!$}, P. L. Garrido{$^@$}}
 
\centerline{\it {$^*$}Matematica, $I^a$ Universit\`a di Roma,
P.le Moro 2, 00185 Roma, Italia}
\centerline{\it {$^!$}Fisica, $I^a$ Universit\`a di Roma,
P.le Moro 2, 00185 Roma, Italia}
\centerline{\it {$^@$}Instituto Carlos I de F{\'\i}sica
Te{\'o}rica y Computacional,
Universidad de Granada}
\centerline{\it E-18071 Granada, Espa\~na}
 
\vskip0.5truecm
{\it Abstract: The chaotic hypothesis discussed in [GC1]  is tested
experimentally in a simple conduction model. Besides a confirmation of
the hypothesis predictions the results suggest the validity of the
hypothesis in the much wider context in which, as the forcing strength
grows, the attractor ceases to be an Anosov system and becomes an Axiom
A attractor. A first test of the new predictions is also attempted.}
\*
 
 
\vskip0.5truecm
\0{\it\S1. Introduction.} \numsec=1\numfor=1\pgn=1
\*
 
A principle holding when motions have an empirically chaotic
nature was introduced in reference [GC2]:
 
\* {\it Chaotic hypothesis: A many particle system in a stationary state
can be regarded as a smooth dynamical system with a
transitive\annota{1}{\nota\rm The notion of transitivity used in the above
reference, and in the related ones, is  that the stable and unstable
manifolds of each point of the attractor are dense on the attractor.}
Axiom A global attractor for the purpose of computing macroscopic
properties. In the reversible case it can be regarded, for the same
purposes, as a smooth transitive Anosov system.}
\*
 
For an informal discussion of the properties of Anosov systems relevant
for this work, in particular for the ``Boltzmanian representation'' of
the SRB distribution possible for them, see [GC2], [G1].  See [AA],
[Sm] for a
general discussion of the basic geometrical ideas and [S],
[R1],[R2], [Bo] for the original and complete descriptions of the
mathematical notion and properties of Anosov and Axiom A systems.
 
The results of this work mainly concern the reversible Anosov case: in the
concluding remarks we discuss various questions related to reversibility
and to the Axiom A cases. Therefore the part of the hypothesis that
concerns reversible systems is essential for our applications. It can be
rephrased in various ways and by doing so one gains some insights into
the meaning of it: see \S6.
 
This implies that the macroscopic time averages are described by a
probability distribution $\m$ on the "phase space" $\CC$ of {\it
observed events}, also called {\it timing events}, (which could be, for
instance, the occurrence of a microscopic binary "collision").  The time
evolution, or the {\it dynamics}, is a map $S$ of $\CC$ into itself. The
map $S$ is derived from the flow $Q_t$ that solves the differential
equations of motion of the system: the timing events $\CC$ have to be
thought as a surface transversal to the flow and if $t(x)$ is the time
between the timing event $x$ and the successive one $Sx$ it is
$Q_{t(x)}x=Sx$. Note that the points $Q_tx$ {\it are not} timing events
(\ie they are not in $\CC$) for the intermediate times $0< t<
t(x)$.\annota{2}{\nota One may wonder why we do not time the observations at
constant pace: this would indeed be possible.  It is however convenient
to time the observations on natural events ({\it i.e.}, using the
jargon, "to make observations on a Poincar\'e's section") so that one
eliminates one degree of freedom as well as the corresponding
(trivially zero) Lyapunov exponent, as recognized very early, [L].} We
call $Q_t$ the {\it continuous time evolution} and $S$ the {\it
discrete} time evolution.
 
The existence of the distribution $\m$ is assumed in general, as stated
by the following (extension) of the zeroth law, [UF], giving a global
property of the motions generated by initial data chosen randomly with
distribution $\m_0$ proportional to the volume measure on $\CC$:
\*
\0{\it Extended zero-th law: A dynamical system $(\CC,S)$ modeling a
many particle system (or a continuum such as a fluid) describes motions
that admit a statistics $\m$ in the sense that, given any (smooth)
macroscopic observable $F$ defined on the points $x$ of the phase space
$\CC$, the time average of $F$ exists for all $\m_0$--randomly--chosen
initial data $x$ and is given by:
$$\lim_{M\to\io}\fra1M\sum_{k=0}^{M-1}
F(S^jx)=\ig_\CC\m(dx')F(x')\,\,{\buildrel def \over =}\,\,
\media{F}_+\Eq(1.1)$$
where $\m$ is a $S$--invariant probability distribution on $\CC$.}
\*
 
The chaotic hypothesis was proposed by Ruelle in the case of fluid
turbulence, and it is extended to non equilibrium many particle systems
in [GC1].  If one assumes it, then it {\it follows} that the zeroth law
holds, [S],[Bo],[R1]; however it is convenient to regard the two statements
as distinct because the hypothesis we make is "{\it only}" that one can
suppose that the system is Anosov for "practical purposes": this leaves
the possibility that it is not strictly speaking such, and some
corrections ("negligible in the thermodynamic limit")  may be needed on
the predictions obtained by using the hypothesis.
 
In [GC2] the generality of the hypothesis is discussed and in [GC1],
[GC2], we derived, as a rather general consequence, predictions testable
at least by numerical or physical experiments in systems with few
degrees of freedom.  The feature of the prediction relevant for
numerical experiments, a {\it large deviation theorem} or {\it
fluctuation theorem}, is that it is {\it parameter free}; other results
concern the Onsager reciprocity in various classes of mechanical systems
[G5], or fluid models [G4].
 
The theory was developed to understand the results in [ECM2] which,
therefore, provided also the {\it first} test. In this paper we present
the results of numerical experiments that we conducted in order to check
the hypothesis in models different from the shear flow model in [ECM2].
 
Being a rather general principle, the chaotic hypothesis yields
predictions that are sharp and {\it inescapable}, without free
parameters. Hence it is important to check it with the highest precision
possible. This immediately leads one to work at the limit of the present
day computer capabilities and to lengthy data elaboration.
 
In \S2 we describe the models and give a quantitative description of
their rough characteristics, discussing the experiments that we perform.
In \S3 we explain, through an analogy with well known Ising model
properties, the mechanism which allows us to acquire {\it exact}
knowledge of some propeties of $\m$ without actually computing $\m$
itself (which might be a surprising ``achievement'' if one does not
examine [GC1],[GC2] in some detail). In \S4 we present the raw
experimental results; in \S5 and in the Appendix we briefly summarize
the methods followed in statistical analysis of the result. In \S6 we
present comments, plans and perspectives, and some challenging pictures
that seem to emerge from [G4],[G5][BG] and from the present paper.

\1
\*
\0{\it\S2. The models and a description of the experiments.}
\numsec=2\numfor=1
\*
 
The models contain thermostat mechanisms in order to enable the systems
to reach a non equilibrium stationary state in the presence of an
imposed external field: therefore they are related to electrical
conductivity problems. They represent a gas of $N$ identical particles
with mass $m$, interacting via a hard core {\it pair potential} $\f$
with radius $r$ and with an external potential $\f^e\ne0$. The gas is
enclosed in a $2$ dimensional box $[-\fra{k}{2}L,\fra{k}{2}L]\times
[-\fra{\ell}{2}L,\fra{\ell}{2}L]$, $k,\ell=1,2,\ldots$, and is subject
to periodic boundary conditions and to a horizontal constant external
field $E\V i$ ($\V i$ is a unit vector in the $x$--direction). The
external potential will also be just a hard core interaction excluding
access to the area covered by some obstacles or forbidding the crossing
of the box walls.  The obstacles are hard disks with centers situated on
two square lattices with spacing ${L}$ shifted by $\fra12L$ relative to
each other.  The radii of the disks of each sub lattice are equal and
respectively given by $R_1$ and $R_2$, so fixed that {\it every}
rectilinear trajectory must suffer collisions with them. An alternative
setting could have been a collection of identical hard disks (with large
radii) with centers on a triangular lattice: the adopted geometry is the
same as that of the previous paper [GG].
 
The geometry is very simple and the position space is described in Fig.1:

\eqfig{300pt}{100pt}{
\ins{147pt}{12pt}{${\nota L}$}
\ins{148pt}{27pt}{${\nota \V v}$}
\ins{115pt}{38pt}{${\nota R_2}$}
\ins{140pt}{62pt}{${\nota R_1}$}
}{x1}{\eq(2.1)}
 
\didascalia{Fig.1: General billiard structure with scatterers of radius
$R_1$ and $R_2$ in a periodic box with side length $L$, (case
$k\times\ell=1\times1$).}

The box $[-\fra{k}2L,\fra{k}2L]\times [-\fra{\ell}{2}L,\fra{\ell}{2}L]$
consists of $k\cdot \ell$ unit lattice cells with side $L$ joined to
form a square box: at the box boundary we impose periodic boundary
conditions ($pbc$) or, alternatively, semi periodic boundary conditions
($\fra12pbc$): in the latter case the "horizontal" box walls are
reflecting. The experiments with different boundary conditions have
been performed ``completely'' independently, on different machines and with
different codes.
 
The system is in contact with a "thermostat" which adds (or subtracts)
energy so that the total internal energy stays rigorously constant.
The equations of motion are:
 
$$\eqalign{&\dot{\V q}_j=\fra1m \V p_j,\qquad \dot{\V p}_j=\V F_j+E\V
i-\a(\V p)\V p_j\cr &\V F_j\=-\sum_{i\ne j}\Dpr_{\V q_j}\f(\V q_j-\V
q_i)-\Dpr_{\V q_j}\f^e(\V q_j)\cr}\Eq(2.2)$$
\0with $j=1,\ldots,N$; $\a(\V p)=E\,\V i\cdot\sum_j \V p_j/\big(\sum_j
\V p_j^2\big)$ and $\V F_j$ is the (impulsive) force acting on particle
$j$ (which is due only to the hard cores).  The $\a$--term incorporates
the coupling to a "Gaussian thermostat" and it is assumed to obey Gauss'
``principle of least constraint", see [LA].  The constraint here is the
constancy of the internal (\ie kinetic) energy:
 
$$H(\V p,\V q)=\sum_{i=1}^N\fra{\V p_i^2}{2m}\,\,{\buildrel def \over
=}\,\, e_0 N\Eq(2.3)$$
 
\0a typical nonholonomic constraint; it follows then from Gauss'
principle that the force corresponding to the constraint is proportional
to the gradient with respect to $\V p_j$ of $H$. This model has been
studied in great detail, in [CELS], in the case $N=1$; a similar model
has been investigated numerically in [BEC], and very recently in [DPH],
[DM]. It is part of a wide class of models, see [GC1],[GC2], whose
interest for the theory of non equilibrium stationary states was pointed
out in [HHP], [PH], where one can find the first studies performed in
the context in which we are interested.
 
The timing events that we choose to follow are simply the collisions:
whether with the walls or with the obstacles or with other
particles. The boundary conditions will be periodic in both directions
or reflecting in the one perpendicular to the field (``vertical'') and
periodic in the other (``horizontal'').
 
The initial data will be fixed by a random choice with absolutely
continuous distribution on the full phase space $\FF$ (\ie on the full
energy surface).
 
The dimension of the phase space $\FF$ of this system is that of the
energy surface, \ie $4N-1$, and that of the set of timing events  $\CC$
is $2D$ with $D=2N-1$, \ie one unit less than the dimension of $\FF$.
{\it The phase space ``contraction'' rate}, \ie the divergence of the
right hand side of \equ(2.2), is $\g(x)=D\a(x)$. It can be written in
the form:
 
$$\g(x)=D\a(x)= \fra{\e(x)}{k_B T(x)+\fra{2N}{2D}
\fra1{2m}\lis {\V p}^2}\Eq(2.4)$$
 
\0where $\lis{\V p}=\fra1N\sum_j{\V p}$ is the average momentum, $\e(x)$
is the work done on the system per unit time by the external field and
$\fra12k_B T(x)$ is $\fra1D\sum_j\fra1{2m}{(\V p_j-{\lis{\V p}})^2}$
which, if $k_B$ is Boltzmann's constant, defines the {\it kinetic
temperature}: hence the name of {\it entropy production rate} per
(kinetic) degree of freedom that will be occasionally given to $\a(x)$.
Note that $\g(x)$ {\it does not have a definite sign}. In dimension
$d\ge2$ the factor $\fra{2N}{2D}$ would become $\fra{2N}{dD}$.
 
The above contraction corresponds to a contraction of the volume in the
full phase space $\FF$, between one collision and the next, given by
$e^{-t_0\s(x)}$ with:
 
$$\s(x)=\fra1{t_0}\ig_0^{t(x)} \g(Q_tx)\,dt\Eq(2.5)$$
 
\0if $Q_t$ is the continuous time evolution (see \S1) and
$t_0$ denotes the {\it mean collision} time $t_0=\media{t(\cdot)}_+$.
 
We shall be concerned mostly with the contraction rate $\s_\t(x)$
occurring during $\t$ time steps as the system evolves between
$S^{-\t/2}x$ and $S^{\t/2}x$:
 
$$\s_\t(x)\DEF\fra1\tau\sum_{j=-\t/2}^{\t/2-1}\s(S^jx)\Eq(2.6)$$
It has been proved, [CELS], that for $N=1$ and small $E\ne0$ the average
$\media\s_+$ is positive, \ie the system is {\it dissipative}. There
seems to be no reason to think that $\media\s_+$ is not positive when
$N>1$ and our
experiments show that indeed this seems to be the case.
 
Recently, in fact, it has been shown that under very general conditions
(essentially under the assumption that the extended zero-th law holds)
it must be $\media{\s}_+\ge0$, [R3]. In the latter paper it is also
shown that $\media{\s}_+>0$ if the SRB distribution $\m$ gives
probability $1$ to a set which has zero Liouville measure (\ie if the
attractor is "really" smaller than the full phase 
space).\annota{3}{\nota We
call {\it attractor} a set $G$ with minimal (Hausdorff) dimension
which has the property that $\m(G)=1$, \ie which has probability $1$
in the stationary state. In general the closure ${\rm clos}(G)$ of the
attractor may be the whole space $\CC$ while the dimension of $G$ may be
much less.}
 
Therefore it is natural to write:
 
$$\s_\t(x)=\media{\s}_+ p(x)\Eq(2.7)$$
 
\0so that the contraction of the phase space volume, while the system
evolves between $S^{-\t/2}x$ and $S^{\t/2}x$, is $e^{-\t t_0\media{\s}_+
p(x)}$ and $\media{p}_+\=1$.
 
In this case the time reversal map $i$ defined by $i:(\V q,\V p)\to (\V
q,-\V p)$ is such that $\s_\t(ix)=-\s_\t(x)$.
 
The number $k\cdot l$ of unit lattice cells forming the box containing
the gas will be called the {\it size} of the box.  We define the {\it
density} as $\r=\fra{N}{k\,\ell\,L^2}$ and the {\it energy density} as
$e_0=\fra1N\sum_j \fra1{2m}\V p^2_j$ and we take units so that
$m=1,e_0=1/2,L=1$. The properties of the system are thus governed by the
values of the parameters $R_1, R_2, r$ and $\r$;
in place of $\r$ one could use the {\it occupied volume} $\d=\fra{N
V_0}{V-V_{obs}}$ where $V_0$ is the particles volume, $V=k\ell L^2$ is the
box volume and $V_{obs}$ is the total volume of the obstacles, so that
$\d$ is the ratio between the volume occupied by the particles cores and
the free volume where they can roam (\ie the volume of the cell outside
the volume $V_{obs}$ occupied by the obstacles).
 
The field intensity $E$ is fixed to $E=1$  in all experiments. We shall
consider systems with $N=2, N=10$, radii $R_1=0.2,\, R_2=0.4$ and
$r=0.005$ or $r=0.01$ depending on the type of boundary conditions
denoted, above, by $pbc$ and $\fra12pbc$.  Here is a list of the above
kinematic quantities in the cases that we shall consider:
 
%\paginazero
\vskip5mm
\halign{#\2             &#\2         &#\2                     \cr
\noalign{\vskip1.mm}
{\it density}           &$\r$        &$ \fra{N}{V}$           \3
{\it energy density}    &$e_0$         &$\fra12$                     \3
{\it mass}              &$m$         &$1$                \3
{\it box side}          &$L$         &$1$                     \3
{\it obstacles radii}   &$R_1,R_2$   &$0.2,\,\,0.4$   \3
{\it particles radii}   &$r$         &$0.005(pbc)$ {\sl or\ }
$0.01(\fra12pbc)$\3
{\it occupied volume}   &$\d$        &$\fra{N V_0}{V-V_{obs}}$\3
{\it size}              &$k\times l$         &$1\times1(pbc){\,\sl or\,}
2\times 2(\fra12pbc,\, N=2){\,\sl or\,}4\times 5(\fra12pbc,\, N=10)$\3
{\it particles number}   &$N$         &$2{\,\sl or\,}10$      \cr}
\vskip5mm
 
The following {\it dynamical} quantities are particularly interesting
for the qualitative picture of the motions.\\
\bottone The {\it average timing} $t_0$ of the collisions, equal to the
future average $\media{t(\cdot)}_+$ of the time $t(x)$ elapsing between
two successive collisions.\\
\bottone The {\it collision rate} $\n$ will be the number of collisions
between moving particles divided by the total number of collisions
including the ones with the obstacles and the walls (the latter are
present only in the $\fra12pbc$ case).\\
\bottone The {\it average entropy creation per collision}, equal to the 
future average $t_0\media{\s}_+$.\\
\bottone The {\it Lyapunov} exponents $\l_{\max}$ and
$\l_{\min}$ defined, respectively, by the largest expansion rates of line
elements under the action of the positive iterates of the evolution map
$S$ or by the minimum absolute value of the expansion or contraction
rates.\\
\bottone The entropy correlation time defined by the decay rate
$\media{\s_{\t}(S^nx)\s_\t(x)}_+=O(e^{-\th n})$  of the  entropy
autocorrelation (recall that $Q_t$ denotes the continuous time
evolution, see \S1).
\*

The data below have been obtained empirically, \ie without any
attempt at estimating errors, and are useful to get an idea of the basic
qualitative properties of the system. For $N=2$ particles with $e_0=1/2$,
$m=L=1$, $E=1$, density, $k,\ell$,  and radii as above:
 
{\halign{#\2&\2#\hfill&#\2&#\2&#\2&#\cr
\noalign{\vskip3.mm}
{ } & &\2$pbc,\ k=1$&\2$\fra12pbc,\ k=2$ & \3
$\r$&\it density&\2  $2$ &\2  $0.5$ & \3
$\d$&\it occup. vol.&\2 $4.23\cdot 10^{-4}$&\2
$4.23\cdot 10^{-4}$ & \3
$\l_{\max}^{-1}$&{} &\2  $1.32$ & \2 $1.31$ &\3
$\l_{\min}^{-1}$&{}& \2 $1.53\cdot 10^1$ 
& \2 $2.16\cdot 10^1$ &&\kern-0.5truecm\eq(2.8)\3
$\th^{-1}$&$\forall \t$& \2 $\le20$ & \2 $\le20$ & \3
$t_0$&\it timing pace&\2  $1.50\cdot10^{-1}$ &\2 $1.35\cdot 10^{-1}$& \3
$\n$&\it collision rate&\2 $1.09\cdot 10^{-2}$ &\2 $4.93\cdot 10^{-3}$& \3
$\fra{t_0}{2N-1}\media{\s}_+$ & $\fra{entropy\ prod.}{deg. freedom}$& \2
$1.75\cdot10^{-2}$ &\2$1.49\cdot 10^{-2}$ &\cr}}
\vskip5mm
 
\0where only the errors (not shown, but  amounting to less than $0.1\%$;
see below) on $t_0, \media{\s}_+$ have been measured with care since we
need them in our experiments. The other data are purely indicative of
the orders of magnitude; $\forall\t$ means for all values of $\t$
considered below. 
%The data marked by $?$ have not been obtained. The
%reciprocal of the Lyapunov exponents are divided by $t_0$ so that they
%are effectively measured in numbers of collisions. The reciprocal of the
%Lyapunov exponents are divided by $t_0$ so that they are effectively
%measured in numbers of collisions.

%\paginazero
For $10$ particle systems:
\vskip3mm
 
\halign{#\2&\2#\hfill&#\2&#\2&#\2&#\cr
\noalign{\vskip3.mm}
   { }&   &\2$pbc,\ k=1$&\2$\fra12pbc,\ k=2$ & \3
$\r$&\it density&\2 $10$ &\2 $0.5$ & \3
$\d$&\it occup. vol.&\2 $2.11\cdot 10^{-3}$&\2 $4.23\cdot 10^{-4}$ & \3
$\l_{\max}^{-1}$&&\2 $3.50$ &\2 $4.71$ & \3
$\l_{\min}^{-1}$&& \2 $3.12\cdot 10^2$ &\2 $2.20\cdot 10^3$ & \3
$\th^{-1}$&$\forall\t$& \2$<20$ & \2 $<20$ &&\kern-0.5truecm\eq(2.9) \3
$t_0$&\it timing pace&\2 $2.87\cdot10^{-2}$&\2$2.8\cdot 10^{-2}$ & \3
$\n$&\it pair collisions&\2 $8.95\cdot 10^{-2}$  &\2
$8.92\cdot 10^{-3}$ & \3
$\fra{t_0}{2N-1}\media{\s}_+$& $\fra{entropy\ prod.}{deg. freedom}$
&\2 $4.07\cdot10^{-3}
$&\2 $3.52\cdot 10^{-3}$ &\cr}
\vskip5mm
\0with the same comments on the errors and the symbols as presented in
\equ(2.8).
 
We shall study the probability distribution $\p_\t(p)\,dp$, {\it in the
stationary state} $\m$, of the variable $p$ that is defined by \equ(2.6)
above for $\t$ large.  In fact the theory of the chaotic hypothesis
foresees that, if $\t$ is large compared to $\l_{\min}^{-1}$, then
$\p_\t(p)$ verifies:
 
$$\log\fra{\p_\t(p)}{\p_\t(-p)}=\t t_0\media{\s}_+\,p\Eq(2.10)$$
 
This is the content of the {\it fluctuation theorem} discussed in
[GC1],[CG2],[G2],[G1] and it means that the {\it odd} part of
$\log\p_\t(p)$ is linear in $p$ with an \ap determined slope. {\it
Nothing} is known about the {\it even} part.
 
One may think that the even part is proportional to $p^2$, \ie
$\log\fra{\p_\t(p)}{\p_\t(1)}=-\fra14(p-1)^2\t t_0\media{\s}_+$;
therefore it is interesting to check whether the {\it kurtosis}:
 
$$\k_\t=\fra{\media{(p-1)^4}_+-3\media{(p-1)^2}^2_+}{\media{(p-1)^2}^2_+}
\Eq(2.11)$$
vanishes, if $\media{\cdot}_+$ denotes average with respect to the SRB
distribution $\m$ in \equ(1.1) (recall that $\k_\t\=0$ for a Gaussian
distribution and the value $\k_\t$ can be taken as a quantitative
dimensionless estimate of the ``non gaussian'' nature of the
distribution).  The central limit theorem for transitive Anosov systems,
[S], implies a Gaussian distribution for the variable $\s_\t$ for large
$\t$; this yields information about the deviations of $\t\s_\t$ from its
average $\t\media{\s}_+$ by quantities of order $\sqrt\t$; but \equ(2.10)
describes properties of {\it large deviations}, proportional to $\tau$.
Therefore there is no {\it a priori} reason to expect that the
$\p_\t(p)$ is Gaussian; hence we {\it do not expect} that $\k_\t=0$; and
the evaluation of $\k_\t$, once \equ(2.10) is established, is of
considerable interest.
 
Note that the variable $\s_\t$ varies on a finite range, at fixed $N$.
This means that $p$ varies in a finite interval $[-p^*,p^*]$, symmetric
by the time reversal symmetry, whose size can be easily measured and an
idea of its value can be obtained from the following rough data for 
$\t=20$:
\*
\halign{$#$&$#$\2&$#$\2&$#$\2&$#$\2&#\2\cr &pbc,\ N=2& \fra12pbc,
N=2&pbc,\ N=10& \fra12pbc, N=10 \cr\noalign{\vskip3.mm} p^*\kern1cm&
9.91& 9.25&7.92&8.55 &\hbox{\kern1truecm}\hfill
\eq(2.12)\cr}
that give the values of $p^*$ on the actually observed trajectories.
The definition of $\a$ and \equ(2.5), by using the Schwartz
inequality, imply a bound:
$$p^*\le \fra{2N-1}{\media{\s}_+}\fra{E}{\sqrt{2 e_0}}\fra{t_{\max}}{t_0}
\Eq(2.13)$$
if $e_0$ is the energy per particle and $t_{\max}$ is the maximum time
between collisions. And the bound is saturated when all the particles 
have the same velocity parallel to the field. 
 
Finally the result \equ(2.10) is valid in the {\it limit} $\t\to\io$ and
this means that, in order to check it, one has to perform many
experiments with various values of $\t$: one expects that $\t$ should be
large compared to $\th^{-1}$, {\it at least at fixed $p$} (but the
errors are not expected to be uniform in $p$ so that one should not be
surprised to see still corrections at large values of $p$ for values of
$\t$ for which \equ(2.10) holds without appreciable corrections at small
values of $p$).
 
In this experiment we have computed the distribution $\p_\t(p)$ at fixed
$\t$ (using the discrete evolution $S$) by measuring $p$ over time
intervals of length $\t$ but {\it spaced by a fixed number of
collisions} $\D$ during which no measurement is made. The time interval
$\D$ has been taken large compared to the relevant characteristic times
of the system, \ie the average free flight time or the inverse of the
time decay constant for the entropy autocorrelation function. The latter
times being of the same order of magnitude and of the order of $1$ to
$10$  collision times, we took $\D=50 $ collisions, see
\equ(2.8),\equ(2.9). Larger $\D$ would have been better: but we would
lose statistics.  We then assume, in the statistical analysis, that the
data so obtained are uncorrelated.
 
We made some empirical tests that this time delay was sufficiently large
by investigating, in various cases, how important the time correlations
were in the evaluation of the errors. Not unexpectedly we found that the
statistical errors decrease by increasing  the sampling delay $\D$, in
spite of the smaller statistical samples: so as far as statistical
errors are concerned there is some (small) advantage in taking
measurements spaced by $\D$ large. But mainly this was a test of our
independence assumption used in the errors theory. If there had been a
drastic change in the statistical errors we would have concluded that
the correlation time for the entropy production was not of the order of
$10$ collision times.
 
Another time scale of interest is $\l_{\min}^{-1}$: this is however very
difficult to measure and it may have value $+\io$ for some values of
$E$, (see Fig.15 below).  There is no evidence that this time scale
is related to the entropy autocorrelation: {\it which appears to be short
ranged}, as far as we can see.  On the other end there is evidence that
as $E$ grows the number of positive Lyapunov exponents decreases.  In
\S6 we try to establish a connection between this observation and the
fluctuation theorem.
 
This is very important for us as it shows that the chaotic hypothesis
may hold in a far stronger sense than originally meant in [GC1], [GC2].
One can see that if the dimension of the stable manifold of the
attractor points is not equal to that of the unstable manifold, then the
closure of attractor for the forward motion cannot be the same as that
of the attractor for the backward motion. Hence the time reversal cannot
leave the attractor invariant {\it but} the chaotic hypotesis, as
formulated in \S1, tells us that {\it nevertheless} the motion {\it on the
attractor} is reversible in the sense that there is a map $i^*$ which
leaves the attractor invariant and changes $S$ into $S^{-1}$. The map
$i^*$ (whose existence is far from obvious) will be called the {\it
local time reversal} on the attractor, see \S6,[BG].
 
\*
\0{\it\S3. Ising model analogy.}
\numsec=3\numfor=1\*
 
The fluctuation theorem as expressed by \equ(2.10) and the subsequent
comments on the Gaussian nature of the function $\p_\t(p)$ may seem
somewhat strange and unfamiliar.
 
It is therefore worth pointing out that the phenomenon of a ``linear
fluctuation law'' on the odd part of the distribution, in the sense of
\equ(2.10), without a globally Gaussian distribution, is in fact well
known in statistical mechanics and probability theory. And an example of
what the fluctuation theorem means in a concrete case, in which
$\p_\t(p)$ is {\it not} Gaussian, can be made by using the Ising model
on a $1$ dimensional lattice $Z$.
 
We consider the space $\CC$ of the {\it spin configurations} $\V
\s=\{\s_\x\}, \,\x\in Z$ and the map $S$ that translates each
configuration to the right (say). The ``time reversal'' is the map
$i:\{\V\s\}\to\{-\V \s\}$ that changes the sign to each spin.
 
The probability distribution that approximates the SRB distribution is
the finite volume Gibbs distribution:
 
$$\m_\L(\V \s)=\fra{\exp\big(J\sum_{j=-T}^{T-1}\s_j\s_{j+1}\,+\,
h\sum_{j=-T}^T \s_j\big)}{normalization} \Eq(3.1)$$
where $\L=[-T,T]$ is a large interval, $J, h>0$. The configuration $\V
\s$ outside $\L$ is distributed independently on the one inside the box
$\L$, to fix the ideas.
 
Calling $\media{m}_+$ the average magnetization in the thermodynamic
limit we define the magnetization in a box $[-\fra\t2,\fra\t2]$ to be
$M_\t=\t \media{m}_+ p$ and we look at the probability distribution
$\p_\t^T(p)$ of $p$ in the limit $T\to\io$. The Gibbs distribution
corresponding to the limit of \equ(3.1) will play the role of the SRB
distribution. Calling this limit probability $\p_\t(p)$ it is easy to
see that:
 
$$\fra{\p_\t(p)}{\p_\t(-p)}\,\tende{\t\to\io}\,e^{2\t h\media{m}_+\,p}
\Eq(3.2)$$
This is in fact obvious if we take the two limits $T\to\io$ and
$\t\to\io$ {\it simultaneously} by setting $T=\fra\t2$. In such a case,
if $\sum_{\V\s;\,p}$ denotes summation over all the configurations with
given magnetization in $[-T,T]$, \ie such that
$\sum_{j=-\fra\t2}^{\fra\t2-1}\s_j=\media{m}_+ \,p$
the distribution \equ(3.1) gives us immediately that:
 
$$
\fra{\p^T_\t(p)}{\p^T_\t(-p)}=
\fra{\sum_{\V\s,\, p}\exp{J\sum_{j=-T}^{T-1}\s_j\s_{j+1}\,+\,
h\sum_{j=-T}^T \s_j}}{\sum_{\V\s,\, -p}\exp{J
\sum_{j=-T}^{T-1}\s_j\s_{j+1}\,+\,h\sum_{j=-T}^T \s_j}}
\=e^{2\t h\media{m}_+\,p}\Eq(3.3)$$
if we use the symmetry of the pair interaction part of the energy under
the ``time reversal'' (\ie under spin reversal).
 
The error involved, in the above argument, in taking $T=\fra\t2$ rather
then first $T\to\io$ and then $\t\to\io$, can be easily corrected since
the corrections are ``boundary terms'', and in one dimensional short
range spin systems there are no phase transitions and the boundary terms
have no influence in the infinite volume limit (\ie they manifest
themselves as corrections that vanish, as $T\to\io$ followed by
$\t\to\io$).
 
One may not like that the operation $i$ commutes with $S$ rather than
transforming it into $S^{-1}$. Another example in which the operation
$i$ does also invert the sign of time is obtained by defining $i$ as
$\{i\V\s\}|_j= -\s_{-j}$: the \equ(3.3) can be derived also by using this
new symmetry operation.
 
The above examples show why there is \ap independence between any
Gaussian property of $\p_\t(p)$ and the fluctuation theorem. The theory
of the fluctuation theorem in [GC1] is in fact {\it based} on the
possibility (discovered in [S]) of representing a chaotic system as a
one dimensional short range system of interacting spins (in general
higher that $\fra12$); and the argument is, actually, very close to the
above one for the Ising model with, however, a rather different time
reversal operation. See [G2] for mathematical details on the boundary
condition question.
 
\*

\1
\0{\it\S4. Experimental results.}
\numsec=4\numfor=1 \*
 
We have made several measurements. Each measurement is quite delicate
and time consuming, hence the reader will probably forgive us for not
having done all the experiments that one finds natural to do. Each
experiment requires several days of CPU time on the computers that we
used (and several months to prepare the final runs). The statistical
errors have been measured by $3$ times the standard deviation, and the
other errors are estimated by following the criteria discussed in [GG]
and resumed in the appendix.  Thus the experiments {\it should be
reproducible within our error bars} if the latter are defined as we
do. Hopefully the data we give can be of use even if one decides (for
mathematical reasons or to test other theoretical ideas) to change the
assignments of the errors. In all the following graphs the error bounds
are always marked although sometimes they may be not visible.
\*
\0{\it4.1. Periodic boundary conditions, $N=2$:}
The values of $R_1,R_2$ are respectively $0.2,0.4$; the particle radius
is $0.005$ and the electric field is fixed $E=1.$, a value that seems to
be quite large (see however \S6). The qualitative data of the resulting
evolution are given in the first column of the table \equ(2.8). Other
experiments at varying field intensity are described in \S6.
 
\*
\0{\it4.1.1 The probability distribution $\p_\t(p)$:} The evolution is
studied over $1.08\cdot10^6$ collisions. In Fig.2 we give the graph of
$\p_\t(p)$ for various $\t$. \vskip0.5truecm
 
%%%%%%%%%%%%%%%%%%%%%%%%%%%%
 
%\input figph2_ps.tex
\dimen2=25pt \advance\dimen2 by 95pt
\dimen3=103pt \advance\dimen3 by 40pt
\dimenfor{\dimen3}{125pt}
\eqfigfor{
\ins{123pt}{0pt}{$\hbox to20pt{\hfill ${\scriptstyle p}$\hfill}$}
\ins{0pt}{\dimen2}{$\hbox to25pt{\hfill ${\scriptstyle \pi(p)}
$\hfill}$}
\ins{71pt}{113pt}{$\hbox to51pt{\hfill ${\scriptstyle \tau=20}
$\hfill}$}
\ins{5pt}{3pt}{$\hbox to 22pt{\hfill${\scriptstyle -6.00}
$\hfill}$}
\ins{28pt}{3pt}{$\hbox to 22pt{\hfill${\scriptstyle -3.00}
$\hfill}$}
\ins{51pt}{3pt}{$\hbox to 22pt{\hfill${\scriptstyle 0.00}
$\hfill}$}
\ins{74pt}{3pt}{$\hbox to 22pt{\hfill${\scriptstyle 3.00}
$\hfill}$}
\ins{96pt}{3pt}{$\hbox to 22pt{\hfill${\scriptstyle 6.00}
$\hfill}$}
\ins{0pt}{21pt}{$\hbox to 20pt{\hfill${\scriptstyle 0.00}
$\hfill}$}
\ins{0pt}{33pt}{$\hbox to 20pt{\hfill${\scriptstyle 0.10}
$\hfill}$}
\ins{0pt}{45pt}{$\hbox to 20pt{\hfill${\scriptstyle 0.20}
$\hfill}$}
\ins{0pt}{57pt}{$\hbox to 20pt{\hfill${\scriptstyle 0.30}
$\hfill}$}
\ins{0pt}{69pt}{$\hbox to 20pt{\hfill${\scriptstyle 0.40}
$\hfill}$}
\ins{0pt}{80pt}{$\hbox to 20pt{\hfill${\scriptstyle 0.50}
$\hfill}$}
\ins{0pt}{92pt}{$\hbox to 20pt{\hfill${\scriptstyle 0.60}
$\hfill}$}
\ins{0pt}{104pt}{$\hbox to 20pt{\hfill${\scriptstyle 0.70}
$\hfill}$}
}{figph20}{\ins{123pt}{0pt}{$\hbox to20pt{\hfill ${\scriptstyle p}
$\hfill}$}
\ins{0pt}{\dimen2}{$\hbox to25pt{\hfill ${\scriptstyle \pi(p)}
$\hfill}$}
\ins{71pt}{113pt}{$\hbox to51pt{\hfill ${\scriptstyle \tau=40}
$\hfill}$}
\ins{5pt}{3pt}{$\hbox to 22pt{\hfill${\scriptstyle -6.00}
$\hfill}$}
\ins{28pt}{3pt}{$\hbox to 22pt{\hfill${\scriptstyle -3.00}
$\hfill}$}
\ins{51pt}{3pt}{$\hbox to 22pt{\hfill${\scriptstyle 0.00}
$\hfill}$}
\ins{74pt}{3pt}{$\hbox to 22pt{\hfill${\scriptstyle 3.00}
$\hfill}$}
\ins{96pt}{3pt}{$\hbox to 22pt{\hfill${\scriptstyle 6.00}
$\hfill}$}
\ins{0pt}{21pt}{$\hbox to 20pt{\hfill${\scriptstyle 0.00}
$\hfill}$}
\ins{0pt}{33pt}{$\hbox to 20pt{\hfill${\scriptstyle 0.10}
$\hfill}$}
\ins{0pt}{45pt}{$\hbox to 20pt{\hfill${\scriptstyle 0.20}
$\hfill}$}
\ins{0pt}{57pt}{$\hbox to 20pt{\hfill${\scriptstyle 0.30}
$\hfill}$}
\ins{0pt}{69pt}{$\hbox to 20pt{\hfill${\scriptstyle 0.40}
$\hfill}$}
\ins{0pt}{80pt}{$\hbox to 20pt{\hfill${\scriptstyle 0.50}
$\hfill}$}
\ins{0pt}{92pt}{$\hbox to 20pt{\hfill${\scriptstyle 0.60}
$\hfill}$}
\ins{0pt}{104pt}{$\hbox to 20pt{\hfill${\scriptstyle 0.70}
$\hfill}$}
}{figph21}{\ins{123pt}{0pt}{$\hbox to20pt{\hfill ${\scriptstyle p}
$\hfill}$}
\ins{0pt}{\dimen2}{$\hbox to25pt{\hfill ${\scriptstyle \pi(p)}
$\hfill}$}
\ins{71pt}{113pt}{$\hbox to51pt{\hfill ${\scriptstyle \tau=60}
$\hfill}$}
\ins{5pt}{3pt}{$\hbox to 22pt{\hfill${\scriptstyle -6.00}
$\hfill}$}
\ins{28pt}{3pt}{$\hbox to 22pt{\hfill${\scriptstyle -3.00}
$\hfill}$}
\ins{51pt}{3pt}{$\hbox to 22pt{\hfill${\scriptstyle 0.00}
$\hfill}$}
\ins{74pt}{3pt}{$\hbox to 22pt{\hfill${\scriptstyle 3.00}
$\hfill}$}
\ins{96pt}{3pt}{$\hbox to 22pt{\hfill${\scriptstyle 6.00}
$\hfill}$}
\ins{0pt}{21pt}{$\hbox to 20pt{\hfill${\scriptstyle 0.00}
$\hfill}$}
\ins{0pt}{33pt}{$\hbox to 20pt{\hfill${\scriptstyle 0.10}
$\hfill}$}
\ins{0pt}{45pt}{$\hbox to 20pt{\hfill${\scriptstyle 0.20}
$\hfill}$}
\ins{0pt}{57pt}{$\hbox to 20pt{\hfill${\scriptstyle 0.30}
$\hfill}$}
\ins{0pt}{69pt}{$\hbox to 20pt{\hfill${\scriptstyle 0.40}
$\hfill}$}
\ins{0pt}{80pt}{$\hbox to 20pt{\hfill${\scriptstyle 0.50}
$\hfill}$}
\ins{0pt}{92pt}{$\hbox to 20pt{\hfill${\scriptstyle 0.60}
$\hfill}$}
\ins{0pt}{104pt}{$\hbox to 20pt{\hfill${\scriptstyle 0.70}
$\hfill}$}
}{figph22}{\ins{123pt}{0pt}{$\hbox to20pt{\hfill ${\scriptstyle p}
$\hfill}$}
\ins{0pt}{\dimen2}{$\hbox to25pt{\hfill ${\scriptstyle \pi(p)}
$\hfill}$}
\ins{71pt}{113pt}{$\hbox to51pt{\hfill ${\scriptstyle \tau=100}
$\hfill}$}
\ins{5pt}{3pt}{$\hbox to 22pt{\hfill${\scriptstyle -6.00}
$\hfill}$}
\ins{28pt}{3pt}{$\hbox to 22pt{\hfill${\scriptstyle -3.00}
$\hfill}$}
\ins{51pt}{3pt}{$\hbox to 22pt{\hfill${\scriptstyle 0.00}
$\hfill}$}
\ins{74pt}{3pt}{$\hbox to 22pt{\hfill${\scriptstyle 3.00}
$\hfill}$}
\ins{96pt}{3pt}{$\hbox to 22pt{\hfill${\scriptstyle 6.00}
$\hfill}$}
\ins{0pt}{21pt}{$\hbox to 20pt{\hfill${\scriptstyle 0.00}
$\hfill}$}
\ins{0pt}{33pt}{$\hbox to 20pt{\hfill${\scriptstyle 0.10}
$\hfill}$}
\ins{0pt}{45pt}{$\hbox to 20pt{\hfill${\scriptstyle 0.20}
$\hfill}$}
\ins{0pt}{57pt}{$\hbox to 20pt{\hfill${\scriptstyle 0.30}
$\hfill}$}
\ins{0pt}{69pt}{$\hbox to 20pt{\hfill${\scriptstyle 0.40}
$\hfill}$}
\ins{0pt}{80pt}{$\hbox to 20pt{\hfill${\scriptstyle 0.50}
$\hfill}$}
\ins{0pt}{92pt}{$\hbox to 20pt{\hfill${\scriptstyle 0.60}
$\hfill}$}
\ins{0pt}{104pt}{$\hbox to 20pt{\hfill${\scriptstyle 0.70}
$\hfill}$}
}{figph23}{\eq(4.1)}
 
%%%%%%%%%%%%%%%%%%%%%%%%%%%%
 
\vskip0.5truecm
\didascalia{Fig.2: The histograms for $\p_\t(p)$ at various values of
the observation time $\t=20,40,60,100$. Each vertical bar is the error
bar centered around the measured point. The dots on the axis mark the
extremes of the interval where the observed data differ form $0$ within
the statistical error.}
 
The error bars are very small particularly for the data at the edge of
the observability interval because the data are very many. But the {\it
relative errors} (not shown) are very small at the center and they are
very large at the edges, of course.
 
We have attributed to each point on the above histograms a statistical
error as explained in \S5 below (essentially they have been supposed
independent variables and they have been given an error of $3$ times the
standard deviation). In fact the analysis of dispersion of each value
shows that the "law of large numbers" is obeyed, and the standard
deviation $m_n(\t){\buildrel def\over=} \media{(p-\media{p}_\t)^n}$,
$n=2$, and the third order deviation, $n=3$, approach $0$ with an
apparent decay given by:
$$\eqalign{
m_2(\t)=&-.005(\pm.002)+39.80(\pm0.05)\fra1\t\cr
m_3(\t)=&-.002(\pm.004)+93.(\pm1.5)\fra1{\t^2}\cr}\Eq(4.2)$$
The error analysis leading to \equ(4.2) follows the same scheme of
[GG]. We can also use here the notion of "goodness" introduced in [GG]
to measure how good a fit is (reproduced in the appendix below), and we
measured the goodness of the above fits.  We do not use the word
"reliability test", and use ``goodness'' instead, to avoid inducing the
reader to think that we rely on some standard error analysis.  The
goodness is $2.45\cdot10^{-3}$ and $4.94\cdot10^{-3}$ respectively.
 
The results are given in the following Fig.3:
 
%%%%%%%%%%%%%%%%%%%%%%
 
%\input figpv2_ps.tex
\dimen2=25pt \advance\dimen2 by 95pt
\dimen3=103pt \advance\dimen3 by 40pt
\dimenfor{\dimen3}{125pt}
\eqfigter{
\ins{123pt}{0pt}{$\hbox to20pt{\hfill ${\scriptstyle \tau^{-1}}$
\hfill}$}
\ins{0pt}{\dimen2}{$\hbox to25pt{\hfill ${\scriptstyle m_2(\tau)}$
\hfill}$}
\ins{14pt}{3pt}{$\hbox to 26pt{\hfill${\scriptstyle 0.00}$
\hfill}$}
\ins{40pt}{3pt}{$\hbox to 26pt{\hfill${\scriptstyle 0.03}$
\hfill}$}
\ins{67pt}{3pt}{$\hbox to 26pt{\hfill${\scriptstyle 0.06}$
\hfill}$}
\ins{93pt}{3pt}{$\hbox to 26pt{\hfill${\scriptstyle 0.09}$
\hfill}$}
\ins{0pt}{21pt}{$\hbox to 20pt{\hfill${\scriptstyle 0}$
\hfill}$}
\ins{0pt}{41pt}{$\hbox to 20pt{\hfill${\scriptstyle 1}$
\hfill}$}
\ins{0pt}{60pt}{$\hbox to 20pt{\hfill${\scriptstyle 2}$
\hfill}$}
\ins{0pt}{80pt}{$\hbox to 20pt{\hfill${\scriptstyle 3}$
\hfill}$}
\ins{0pt}{100pt}{$\hbox to 20pt{\hfill${\scriptstyle 4}$
\hfill}$}
}{figpv20}{\ins{123pt}{0pt}{$\hbox to20pt{\hfill ${\scriptstyle \tau^{-2}}$
\hfill}$}
\ins{0pt}{\dimen2}{$\hbox to25pt{\hfill ${\scriptstyle m_3(\tau)}$
\hfill}$}
\ins{7pt}{3pt}{$\hbox to 43pt{\hfill${\scriptstyle 0.000}$
\hfill}$}
\ins{50pt}{3pt}{$\hbox to 43pt{\hfill${\scriptstyle 0.005}$
\hfill}$}
\ins{93pt}{3pt}{$\hbox to 43pt{\hfill${\scriptstyle 0.010}$
\hfill}$}
\ins{0pt}{22pt}{$\hbox to 20pt{\hfill${\scriptstyle 0.0}$
\hfill}$}
\ins{0pt}{35pt}{$\hbox to 20pt{\hfill${\scriptstyle 0.2}$
\hfill}$}
\ins{0pt}{48pt}{$\hbox to 20pt{\hfill${\scriptstyle 0.4}$
\hfill}$}
\ins{0pt}{62pt}{$\hbox to 20pt{\hfill${\scriptstyle 0.6}$
\hfill}$}
\ins{0pt}{75pt}{$\hbox to 20pt{\hfill${\scriptstyle 0.8}$
\hfill}$}
\ins{0pt}{88pt}{$\hbox to 20pt{\hfill${\scriptstyle 1.0}$
\hfill}$}
\ins{0pt}{102pt}{$\hbox to 20pt{\hfill${\scriptstyle 1.2}$
\hfill}$}
}{figpv21}{\ins{123pt}{0pt}{$\hbox to20pt{\hfill ${\scriptstyle \tau^{-1}}$
\hfill}$}
\ins{0pt}{\dimen2}{$\hbox to25pt{\hfill ${\scriptstyle \k(\tau)}$
\hfill}$}
\ins{14pt}{3pt}{$\hbox to 26pt{\hfill${\scriptstyle 0.00}$
\hfill}$}
\ins{40pt}{3pt}{$\hbox to 26pt{\hfill${\scriptstyle 0.03}$
\hfill}$}
\ins{67pt}{3pt}{$\hbox to 26pt{\hfill${\scriptstyle 0.06}$
\hfill}$}
\ins{93pt}{3pt}{$\hbox to 26pt{\hfill${\scriptstyle 0.09}$
\hfill}$}
\ins{0pt}{15pt}{$\hbox to 20pt{\hfill${\scriptstyle -0.6}$
\hfill}$}
\ins{0pt}{28pt}{$\hbox to 20pt{\hfill${\scriptstyle -0.4}$
\hfill}$}
\ins{0pt}{42pt}{$\hbox to 20pt{\hfill${\scriptstyle -0.2}$
\hfill}$}
\ins{0pt}{55pt}{$\hbox to 20pt{\hfill${\scriptstyle -0.0}$
\hfill}$}
\ins{0pt}{69pt}{$\hbox to 20pt{\hfill${\scriptstyle 0.2}$
\hfill}$}
\ins{0pt}{82pt}{$\hbox to 20pt{\hfill${\scriptstyle 0.4}$
\hfill}$}
\ins{0pt}{95pt}{$\hbox to 20pt{\hfill${\scriptstyle 0.6}$
\hfill}$}
}{figpv22}{\eq(4.3)}
 
\didascalia{Fig.3: The decay of the $2^d$ and $3^d$ moments, as
functions of $\fra1\t$ or $\fra1{\t^2}$, and the kurtosis.}
 
The question of whether the probability distribution $\p_\t(p)$ is
Gaussian is investigated by computing the {\it kurtosis} as a function
of $\t$: the latter quantity is a dimensionless parameter and it is
related to the fourth moment:
$\fra{\media{\s_\t^4}-3\media{\s_\t^2}^2}{m_2(\t)}$. It can be fitted by
a law:
$$\k(\t)=[0.00(\pm0.01)+2.3(\pm0.3)\fra1\t]\times\fra1{m_2(\t)^2}
\Eq(4.4)$$
\1
The data are reported in Fig.3.  But, unlike the cases of $m_2(\t)$ and
$m_3(\t)$ the goodness of this fit is $1.63\cdot10^{-2}$ and it is {\it
comparable} with the goodness of fits with a law $\fra1{\t^2}$ or even
$\fra1{\t^3}$: the errors are too big so that many fits are ``as good''
(and also very good: this only shows that the notion of goodness has
shortcomings as much as any other accuracy test).
 
We conclude that the distribution seems compatible with a
Gaussian. Of course one expects a Gaussian behaviour for the small
deviations, \ie for $(p-1)\sim \t^{-1/2}$:
If one assumes that the system is Anosov (or just that it has an Axiom
A attractor) then this follows, as a theorem, from the results of
[S] (or [R2]).
 
However we know that it cannot be Gaussian beyond the range
$O(\t^{-1/2})$, because it has support between $\pm p^*$ with
$p^*<+\io$, see \equ(2.12) and \S3.  Hence we the apparent
closeness to a Gaussian distribution might be an accident, that might
disappear as $\t$ increases. If it does not then this is an interesting
question to examine, see \S6.
 
In Fig.4 we show the main quantity of interest here, \ie the graph of:
$$x(p)=\fra1{\t t_0\media{\s}_+}\log\fra{\p_\t(p)}{\p_\t(-p)}\Eq(4.5)$$
versus $p$ (recall that $\media{\s}_+$ is $(2N-1)j$ if $j$ is the
electric current):
 
%%%%%%%%%%%%%%%%%%%%%%%%%%%
 
%\input figpr2_ps.tex
\dimen2=25pt \advance\dimen2 by 95pt
\dimen3=103pt \advance\dimen3 by 40pt
\dimenfor{\dimen3}{125pt}
\eqfigfor{
\ins{123pt}{0pt}{$\hbox to20pt{\hfill ${\scriptstyle p}$
\hfill}$}
\ins{0pt}{\dimen2}{$\hbox to25pt{\hfill ${\scriptstyle x(p)}$
\hfill}$}
\ins{71pt}{113pt}{$\hbox to51pt{\hfill ${\scriptstyle \tau=20}$
\hfill}$}
\ins{17pt}{3pt}{$\hbox to 19pt{\hfill${\scriptstyle 0}$
\hfill}$}
\ins{37pt}{3pt}{$\hbox to 19pt{\hfill${\scriptstyle 1}$
\hfill}$}
\ins{56pt}{3pt}{$\hbox to 19pt{\hfill${\scriptstyle 2}$
\hfill}$}
\ins{76pt}{3pt}{$\hbox to 19pt{\hfill${\scriptstyle 3}$
\hfill}$}
\ins{96pt}{3pt}{$\hbox to 19pt{\hfill${\scriptstyle 4}$
\hfill}$}
\ins{0pt}{22pt}{$\hbox to 20pt{\hfill${\scriptstyle 0}$
\hfill}$}
\ins{0pt}{36pt}{$\hbox to 20pt{\hfill${\scriptstyle 1}$
\hfill}$}
\ins{0pt}{51pt}{$\hbox to 20pt{\hfill${\scriptstyle 2}$
\hfill}$}
\ins{0pt}{66pt}{$\hbox to 20pt{\hfill${\scriptstyle 3}$
\hfill}$}
\ins{0pt}{81pt}{$\hbox to 20pt{\hfill${\scriptstyle 4}$
\hfill}$}
\ins{0pt}{96pt}{$\hbox to 20pt{\hfill${\scriptstyle 5}$
\hfill}$}
}{figpr20}{\ins{123pt}{0pt}{$\hbox to20pt{\hfill ${\scriptstyle p}$
\hfill}$}
\ins{0pt}{\dimen2}{$\hbox to25pt{\hfill ${\scriptstyle x(p)}$
\hfill}$}
\ins{71pt}{113pt}{$\hbox to51pt{\hfill ${\scriptstyle \tau=40}$
\hfill}$}
\ins{18pt}{3pt}{$\hbox to 16pt{\hfill${\scriptstyle 0.0}$
\hfill}$}
\ins{35pt}{3pt}{$\hbox to 16pt{\hfill${\scriptstyle 0.5}$
\hfill}$}
\ins{51pt}{3pt}{$\hbox to 16pt{\hfill${\scriptstyle 1.0}$
\hfill}$}
\ins{67pt}{3pt}{$\hbox to 16pt{\hfill${\scriptstyle 1.5}$
\hfill}$}
\ins{84pt}{3pt}{$\hbox to 16pt{\hfill${\scriptstyle 2.0}$
\hfill}$}
\ins{100pt}{3pt}{$\hbox to 16pt{\hfill${\scriptstyle 2.5}$
\hfill}$}
\ins{0pt}{21pt}{$\hbox to 20pt{\hfill${\scriptstyle 0}$
\hfill}$}
\ins{0pt}{40pt}{$\hbox to 20pt{\hfill${\scriptstyle 1}$
\hfill}$}
\ins{0pt}{59pt}{$\hbox to 20pt{\hfill${\scriptstyle 2}$
\hfill}$}
\ins{0pt}{78pt}{$\hbox to 20pt{\hfill${\scriptstyle 3}$
\hfill}$}
\ins{0pt}{96pt}{$\hbox to 20pt{\hfill${\scriptstyle 4}$
\hfill}$}
}{figpr21}{\ins{123pt}{0pt}{$\hbox to20pt{\hfill ${\scriptstyle p}$
\hfill}$}
\ins{0pt}{\dimen2}{$\hbox to25pt{\hfill ${\scriptstyle x(p)}$
\hfill}$}
\ins{71pt}{113pt}{$\hbox to51pt{\hfill ${\scriptstyle \tau=60}$
\hfill}$}
\ins{14pt}{3pt}{$\hbox to 22pt{\hfill${\scriptstyle 0.0}$
\hfill}$}
\ins{37pt}{3pt}{$\hbox to 22pt{\hfill${\scriptstyle 0.5}$
\hfill}$}
\ins{60pt}{3pt}{$\hbox to 22pt{\hfill${\scriptstyle 1.0}$
\hfill}$}
\ins{83pt}{3pt}{$\hbox to 22pt{\hfill${\scriptstyle 1.5}$
\hfill}$}
\ins{106pt}{3pt}{$\hbox to 22pt{\hfill${\scriptstyle 2.0}$
\hfill}$}
\ins{0pt}{21pt}{$\hbox to 20pt{\hfill${\scriptstyle 0.0}$
\hfill}$}
\ins{0pt}{38pt}{$\hbox to 20pt{\hfill${\scriptstyle 0.5}$
\hfill}$}
\ins{0pt}{56pt}{$\hbox to 20pt{\hfill${\scriptstyle 1.0}$
\hfill}$}
\ins{0pt}{74pt}{$\hbox to 20pt{\hfill${\scriptstyle 1.5}$
\hfill}$}
\ins{0pt}{91pt}{$\hbox to 20pt{\hfill${\scriptstyle 2.0}$
\hfill}$}
}{figpr22}{\ins{123pt}{0pt}{$\hbox to20pt{\hfill ${\scriptstyle p}$
\hfill}$}
\ins{0pt}{\dimen2}{$\hbox to25pt{\hfill ${\scriptstyle x(p)}$
\hfill}$}
\ins{71pt}{113pt}{$\hbox to51pt{\hfill ${\scriptstyle \tau=100}$
\hfill}$}
\ins{7pt}{3pt}{$\hbox to 34pt{\hfill${\scriptstyle 0.0}$
\hfill}$}
\ins{42pt}{3pt}{$\hbox to 34pt{\hfill${\scriptstyle 0.5}$
\hfill}$}
\ins{76pt}{3pt}{$\hbox to 34pt{\hfill${\scriptstyle 1.0}$
\hfill}$}
\ins{0pt}{20pt}{$\hbox to 20pt{\hfill${\scriptstyle 0.0}$
\hfill}$}
\ins{0pt}{42pt}{$\hbox to 20pt{\hfill${\scriptstyle 0.5}$
\hfill}$}
\ins{0pt}{65pt}{$\hbox to 20pt{\hfill${\scriptstyle 1.0}$
\hfill}$}
\ins{0pt}{87pt}{$\hbox to 20pt{\hfill${\scriptstyle 1.5}$
\hfill}$}
}{figpr23}{\eq(4.6)}
 
%%%%%%%%%%%%%%%%%%%%%%%%%%
\*
\didascalia{Fig.4: The fluctuation theorem test for
$\t=20,40,60,100$. The dashed straight line is the theoretical
prediction of the fluctuation theorem. The arrows mark the point at
distance $\sqrt{\media{(p-1)^2}}$ from $1$. The error bars are inherited
from the histogram data in Fig.2.}
 
The abscissae axis is discretized and the number of events in which $p$
falls in a given interval is taken as proportional to $\p_\t(p)$.
Therefore Fig.4 is a histogram.
 
All values corresponding to different $\t$
collapse on a single straight line in the interval $p\in[0,1.5]$ in the
worst cases (\ie largest $\t$).  For higher values of $p$ (depending on
$\t$) the statistics becomes gradually more and more poor as the
deviations from the mean value of $p$ become too large.
\*

\1
\0{\bf4.1.2.  Conclusions:} From the above data we infer that one cannot
exclude that the distribution is Gaussian.  {\it Assuming} that it is
Gaussian then the fluctuation theorem {\it predicts} a standard
deviation $m_2(\t)=\fra2{\t t_0 \media{\s}_+}\=\fra{A}\t$, see the lines
preceding \equ(2.11) and the above experimental data give:
$$\eqalign{
A=&37.96\pm.21\quad({\rm Gaussian\ assumption})\cr
A=&39.80\pm.05\quad({\rm experiment})\cr}\Eq(4.7)$$
where the Gaussian value is computed from the experimentally measured
$\media{\s}_+$ and $t_0$ (to which we attribute a statistical error of
$3$ times the standard deviation), while the experimental value is
computed from $m_2(\t)$, see \equ(4.2); the error on the second line is
as implied by \equ(4.2).
 
The standard deviations of $p$ from the value $1$ are for
$\t=20,\,40,\,60,\,80,\,100$ respectively $1.41$, $0.99$, $0.81$,
$0.70$ and $0.63$. Such values, multiplied by $3$ (recall that our
conventional statistical errors are $3$ times the standard deviation),
can be arbitrarily assumed to be the boundary between the small and the
large deviations.
 
Note that, as already mentioned in \S2, if the chaotic hypothesis is
assumed then we know from the theory of Sinai that the small deviations
obey a Gaussian law as $\t\to\io$: but the theory in general does not
predict the standard deviation. The example of \S3 is a good
illustration, we believe, of the situation: in that case too we have a
good understanding of the odd part of the distribution, but no grip on
the even part and no reason to know \ap the distribution of the
magnetization (\ie also its even part).  The results may mean that the
values of $\t$ that we reach are not large enough to test the validity
of the central limit theorem. On the other hand the fluctuation theorem
that follows from the chaotic hypothesis is {\it independent} on the
central limit theorem (as the example in \S3 suggests and
illustrates) and therefore the above results are, in our opinion, a good
test of the chaotic assumption. We shall examine this point in more
detail in \S6.
 
Of course the choice, mentioned above, of $3$ standard deviations to
measure the statistical errors and the large deviations "threshold" is
arbitrary. One could, as very often done, decide to use $1$ standard
deviation instead.  Then we could say that we can go quite far in the
large deviation region, but on the other hand many error bars become too
small to be compatible with the data. This is a well known problem with
all experiments and we can just report that it appears also in the
present experiment. It could only be solved by better experiments: hence
we decided to perform experiments in which the computing facilities
available to us were pushed further.  The results will be described
below (when discussing the case of semi-periodic boundary conditions and
$N=2,10$).
\*
 
\0{\bf\S4.2 Periodic boundary conditions, $N=10$:} The geometry is the
same as in the previous case.  The values of $R_1,R_2$ are respectively
$0.2,0.4$; the particle radius is $0.005$ and the electric field is
fixed $E=1.$.  The qualitative data of the resulting evolution are
given in the first column of the table \equ(2.9).
 
\0{\it4.2.1 The probability distribution $\p_\t(p)$:} The evolution is
studied over $7.57\cdot10^7$ collisions.  In Fig.5 we give the graph
of $\p_\t(p)$ for various $\t$.

\1
\dimen2=25pt \advance\dimen2 by 95pt
\dimen3=103pt \advance\dimen3 by 40pt
\dimenfor{\dimen3}{125pt}
\eqfigfor{
\ins{123pt}{0pt}{$\hbox to20pt{\hfill ${\scriptstyle p}$
\hfill}$}
\ins{0pt}{\dimen2}{$\hbox to25pt{\hfill ${\scriptstyle \pi(p)}$
\hfill}$}
\ins{71pt}{113pt}{$\hbox to51pt{\hfill ${\scriptstyle \tau=20}$
\hfill}$}
\ins{11pt}{3pt}{$\hbox to 19pt{\hfill${\scriptstyle -4}$
\hfill}$}
\ins{30pt}{3pt}{$\hbox to 19pt{\hfill${\scriptstyle -2}$
\hfill}$}
\ins{49pt}{3pt}{$\hbox to 19pt{\hfill${\scriptstyle 0}$
\hfill}$}
\ins{68pt}{3pt}{$\hbox to 19pt{\hfill${\scriptstyle 2}$
\hfill}$}
\ins{87pt}{3pt}{$\hbox to 19pt{\hfill${\scriptstyle 4}$
\hfill}$}
\ins{107pt}{3pt}{$\hbox to 19pt{\hfill${\scriptstyle 6}$
\hfill}$}
\ins{0pt}{21pt}{$\hbox to 20pt{\hfill${\scriptstyle 0.0}$
\hfill}$}
\ins{0pt}{31pt}{$\hbox to 20pt{\hfill${\scriptstyle 0.1}$
\hfill}$}
\ins{0pt}{42pt}{$\hbox to 20pt{\hfill${\scriptstyle 0.2}$
\hfill}$}
\ins{0pt}{52pt}{$\hbox to 20pt{\hfill${\scriptstyle 0.3}$
\hfill}$}
\ins{0pt}{62pt}{$\hbox to 20pt{\hfill${\scriptstyle 0.4}$
\hfill}$}
\ins{0pt}{72pt}{$\hbox to 20pt{\hfill${\scriptstyle 0.5}$
\hfill}$}
\ins{0pt}{82pt}{$\hbox to 20pt{\hfill${\scriptstyle 0.6}$
\hfill}$}
\ins{0pt}{92pt}{$\hbox to 20pt{\hfill${\scriptstyle 0.7}$
\hfill}$}
\ins{0pt}{102pt}{$\hbox to 20pt{\hfill${\scriptstyle 0.8}$
\hfill}$}
}{figph100}{\ins{123pt}{0pt}{$\hbox to20pt{\hfill ${\scriptstyle p}$
\hfill}$}
\ins{0pt}{\dimen2}{$\hbox to25pt{\hfill ${\scriptstyle \pi(p)}$
\hfill}$}
\ins{71pt}{113pt}{$\hbox to51pt{\hfill ${\scriptstyle \tau=40}$
\hfill}$}
\ins{11pt}{3pt}{$\hbox to 19pt{\hfill${\scriptstyle -4}$
\hfill}$}
\ins{30pt}{3pt}{$\hbox to 19pt{\hfill${\scriptstyle -2}$
\hfill}$}
\ins{49pt}{3pt}{$\hbox to 19pt{\hfill${\scriptstyle 0}$
\hfill}$}
\ins{68pt}{3pt}{$\hbox to 19pt{\hfill${\scriptstyle 2}$
\hfill}$}
\ins{87pt}{3pt}{$\hbox to 19pt{\hfill${\scriptstyle 4}$
\hfill}$}
\ins{107pt}{3pt}{$\hbox to 19pt{\hfill${\scriptstyle 6}$
\hfill}$}
\ins{0pt}{21pt}{$\hbox to 20pt{\hfill${\scriptstyle 0.0}$
\hfill}$}
\ins{0pt}{31pt}{$\hbox to 20pt{\hfill${\scriptstyle 0.1}$
\hfill}$}
\ins{0pt}{42pt}{$\hbox to 20pt{\hfill${\scriptstyle 0.2}$
\hfill}$}
\ins{0pt}{52pt}{$\hbox to 20pt{\hfill${\scriptstyle 0.3}$
\hfill}$}
\ins{0pt}{62pt}{$\hbox to 20pt{\hfill${\scriptstyle 0.4}$
\hfill}$}
\ins{0pt}{72pt}{$\hbox to 20pt{\hfill${\scriptstyle 0.5}$
\hfill}$}
\ins{0pt}{82pt}{$\hbox to 20pt{\hfill${\scriptstyle 0.6}$
\hfill}$}
\ins{0pt}{92pt}{$\hbox to 20pt{\hfill${\scriptstyle 0.7}$
\hfill}$}
\ins{0pt}{102pt}{$\hbox to 20pt{\hfill${\scriptstyle 0.8}$
\hfill}$}
}{figph101}{\ins{123pt}{0pt}{$\hbox to20pt{\hfill ${\scriptstyle p}$
\hfill}$}
\ins{0pt}{\dimen2}{$\hbox to25pt{\hfill ${\scriptstyle \pi(p)}$
\hfill}$}
\ins{71pt}{113pt}{$\hbox to51pt{\hfill ${\scriptstyle \tau=60}$
\hfill}$}
\ins{11pt}{3pt}{$\hbox to 19pt{\hfill${\scriptstyle -4}$
\hfill}$}
\ins{30pt}{3pt}{$\hbox to 19pt{\hfill${\scriptstyle -2}$
\hfill}$}
\ins{49pt}{3pt}{$\hbox to 19pt{\hfill${\scriptstyle 0}$
\hfill}$}
\ins{68pt}{3pt}{$\hbox to 19pt{\hfill${\scriptstyle 2}$
\hfill}$}
\ins{87pt}{3pt}{$\hbox to 19pt{\hfill${\scriptstyle 4}$
\hfill}$}
\ins{107pt}{3pt}{$\hbox to 19pt{\hfill${\scriptstyle 6}$
\hfill}$}
\ins{0pt}{21pt}{$\hbox to 20pt{\hfill${\scriptstyle 0.0}$
\hfill}$}
\ins{0pt}{31pt}{$\hbox to 20pt{\hfill${\scriptstyle 0.1}$
\hfill}$}
\ins{0pt}{42pt}{$\hbox to 20pt{\hfill${\scriptstyle 0.2}$
\hfill}$}
\ins{0pt}{52pt}{$\hbox to 20pt{\hfill${\scriptstyle 0.3}$
\hfill}$}
\ins{0pt}{62pt}{$\hbox to 20pt{\hfill${\scriptstyle 0.4}$
\hfill}$}
\ins{0pt}{72pt}{$\hbox to 20pt{\hfill${\scriptstyle 0.5}$
\hfill}$}
\ins{0pt}{82pt}{$\hbox to 20pt{\hfill${\scriptstyle 0.6}$
\hfill}$}
\ins{0pt}{92pt}{$\hbox to 20pt{\hfill${\scriptstyle 0.7}$
\hfill}$}
\ins{0pt}{102pt}{$\hbox to 20pt{\hfill${\scriptstyle 0.8}$
\hfill}$}
}{figph102}{\ins{123pt}{0pt}{$\hbox to20pt{\hfill ${\scriptstyle p}$
\hfill}$}
\ins{0pt}{\dimen2}{$\hbox to25pt{\hfill ${\scriptstyle \pi(p)}$
\hfill}$}
\ins{71pt}{113pt}{$\hbox to51pt{\hfill ${\scriptstyle \tau=100}$
\hfill}$}
\ins{11pt}{3pt}{$\hbox to 19pt{\hfill${\scriptstyle -4}$
\hfill}$}
\ins{30pt}{3pt}{$\hbox to 19pt{\hfill${\scriptstyle -2}$
\hfill}$}
\ins{49pt}{3pt}{$\hbox to 19pt{\hfill${\scriptstyle 0}$
\hfill}$}
\ins{68pt}{3pt}{$\hbox to 19pt{\hfill${\scriptstyle 2}$
\hfill}$}
\ins{87pt}{3pt}{$\hbox to 19pt{\hfill${\scriptstyle 4}$
\hfill}$}
\ins{107pt}{3pt}{$\hbox to 19pt{\hfill${\scriptstyle 6}$
\hfill}$}
\ins{0pt}{21pt}{$\hbox to 20pt{\hfill${\scriptstyle 0.0}$
\hfill}$}
\ins{0pt}{31pt}{$\hbox to 20pt{\hfill${\scriptstyle 0.1}$
\hfill}$}
\ins{0pt}{42pt}{$\hbox to 20pt{\hfill${\scriptstyle 0.2}$
\hfill}$}
\ins{0pt}{52pt}{$\hbox to 20pt{\hfill${\scriptstyle 0.3}$
\hfill}$}
\ins{0pt}{62pt}{$\hbox to 20pt{\hfill${\scriptstyle 0.4}$
\hfill}$}
\ins{0pt}{72pt}{$\hbox to 20pt{\hfill${\scriptstyle 0.5}$
\hfill}$}
\ins{0pt}{82pt}{$\hbox to 20pt{\hfill${\scriptstyle 0.6}$
\hfill}$}
\ins{0pt}{92pt}{$\hbox to 20pt{\hfill${\scriptstyle 0.7}$
\hfill}$}
\ins{0pt}{102pt}{$\hbox to 20pt{\hfill${\scriptstyle 0.8}$
\hfill}$}
}{figph103}{\eq(4.8)}
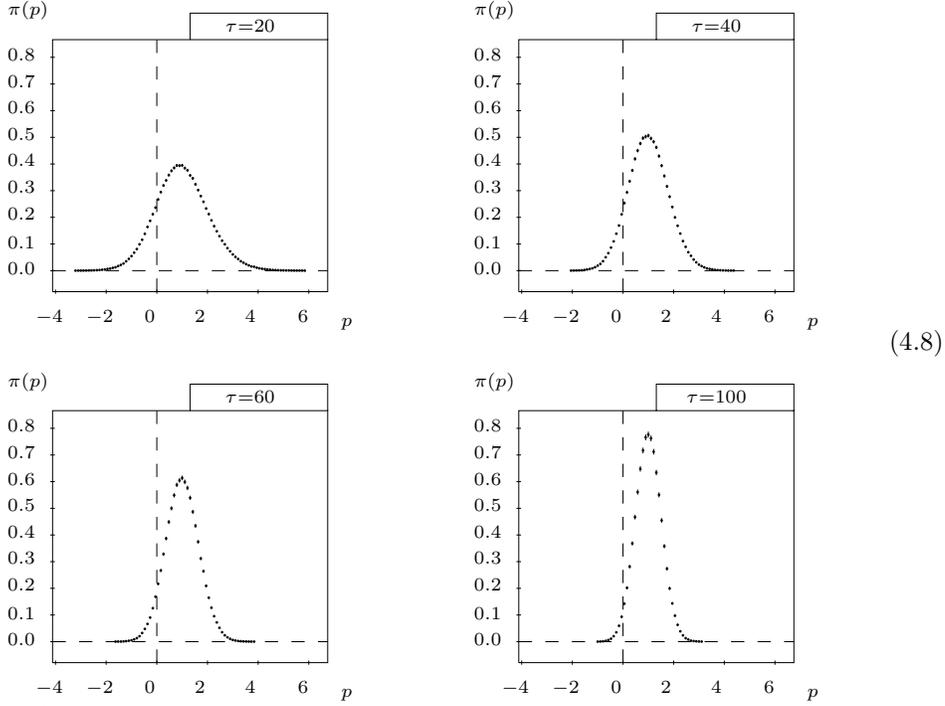
\didascalia{Fig.5: The distribution $\p_\t(p)$ for
$\t=20,40,60,100$.}

\kern-3truemm
\0and in Fig.6 we give the standard deviation and
the third order deviation, as in the previous case as functions of
$\t^{-1}$ and $\t^{-2}$ respectively.

\dimen2=25pt \advance\dimen2 by 95pt
\dimen3=103pt \advance\dimen3 by 40pt
\dimenfor{\dimen3}{125pt}
\eqfigter{
\ins{123pt}{0pt}{$\hbox to20pt{\hfill ${\scriptstyle \tau^{-1}}$
\hfill}$}
\ins{0pt}{\dimen2}{$\hbox to25pt{\hfill ${\scriptstyle m_2(\tau)}$
\hfill}$}
\ins{14pt}{3pt}{$\hbox to 26pt{\hfill${\scriptstyle 0.00}$
\hfill}$}
\ins{40pt}{3pt}{$\hbox to 26pt{\hfill${\scriptstyle 0.03}$
\hfill}$}
\ins{67pt}{3pt}{$\hbox to 26pt{\hfill${\scriptstyle 0.06}$
\hfill}$}
\ins{93pt}{3pt}{$\hbox to 26pt{\hfill${\scriptstyle 0.09}$
\hfill}$}
\ins{0pt}{20pt}{$\hbox to 20pt{\hfill${\scriptstyle 0.0}$
\hfill}$}
\ins{0pt}{39pt}{$\hbox to 20pt{\hfill${\scriptstyle 0.4}$
\hfill}$}
\ins{0pt}{58pt}{$\hbox to 20pt{\hfill${\scriptstyle 0.8}$
\hfill}$}
\ins{0pt}{77pt}{$\hbox to 20pt{\hfill${\scriptstyle 1.2}$
\hfill}$}
\ins{0pt}{96pt}{$\hbox to 20pt{\hfill${\scriptstyle 1.6}$
\hfill}$}
}{figpv100}{\ins{123pt}{0pt}{$\hbox to20pt{\hfill ${\scriptstyle \tau^{-2}}$
\hfill}$}
\ins{0pt}{\dimen2}{$\hbox to25pt{\hfill ${\scriptstyle m_3(\tau)}$
\hfill}$}
\ins{7pt}{3pt}{$\hbox to 43pt{\hfill${\scriptstyle 0.000}$
\hfill}$}
\ins{50pt}{3pt}{$\hbox to 43pt{\hfill${\scriptstyle 0.005}$
\hfill}$}
\ins{93pt}{3pt}{$\hbox to 43pt{\hfill${\scriptstyle 0.010}$
\hfill}$}
\ins{0pt}{21pt}{$\hbox to 20pt{\hfill${\scriptstyle 0.0}$
\hfill}$}
\ins{0pt}{42pt}{$\hbox to 20pt{\hfill${\scriptstyle 0.2}$
\hfill}$}
\ins{0pt}{62pt}{$\hbox to 20pt{\hfill${\scriptstyle 0.4}$
\hfill}$}
\ins{0pt}{82pt}{$\hbox to 20pt{\hfill${\scriptstyle 0.6}$
\hfill}$}
\ins{0pt}{102pt}{$\hbox to 20pt{\hfill${\scriptstyle 0.8}$
\hfill}$}
}{figpv101}{\ins{123pt}{0pt}{$\hbox to20pt{\hfill ${\scriptstyle \tau^{-1}}$
\hfill}$}
\ins{0pt}{\dimen2}{$\hbox to25pt{\hfill ${\scriptstyle \k(\tau)}$
\hfill}$}
\ins{13pt}{3pt}{$\hbox to 26pt{\hfill${\scriptstyle 0.00}$
\hfill}$}
\ins{40pt}{3pt}{$\hbox to 26pt{\hfill${\scriptstyle 0.03}$
\hfill}$}
\ins{66pt}{3pt}{$\hbox to 26pt{\hfill${\scriptstyle 0.06}$
\hfill}$}
\ins{92pt}{3pt}{$\hbox to 26pt{\hfill${\scriptstyle 0.09}$
\hfill}$}
\ins{0pt}{14pt}{$\hbox to 20pt{\hfill${\scriptstyle -0.2}$
\hfill}$}
\ins{0pt}{30pt}{$\hbox to 20pt{\hfill${\scriptstyle 0.0}$
\hfill}$}
\ins{0pt}{46pt}{$\hbox to 20pt{\hfill${\scriptstyle 0.2}$
\hfill}$}
\ins{0pt}{63pt}{$\hbox to 20pt{\hfill${\scriptstyle 0.4}$
\hfill}$}
\ins{0pt}{79pt}{$\hbox to 20pt{\hfill${\scriptstyle 0.6}$
\hfill}$}
\ins{0pt}{96pt}{$\hbox to 20pt{\hfill${\scriptstyle 0.8}$
\hfill}$}
}{figpv102}{\eq(4.9)}
\didascalia{Fig.6: The decay of the $2^d$ and $3^d$ moments, as
functions of $\fra1\t$ or $\fra1{\t^2}$ and of the kurtosis as a
function of $\fra1{\t}$.}

\1 
The fits give:
$$\eqalign{
m_2(\t)=&0.009(\pm.005)+25.24(\pm0.4)\fra1\t\cr
m_3(\t)=&-.0002(\pm.0003)+54.3(\pm1.5)\fra1{\t^2}\cr}\Eq(4.10)$$
and the goodness of the above fits is $1.2\cdot10^{-3}$
and $3.4\cdot10^{-4}$ respectively, for the data with $\t>25$. The data
with $\t<25$ deviate from the above law and we interpret this as finite
size effects (expected from the theory but absent in the previous case
already for $\t>25$).
 
In Fig.6 the kurtosis graph is reported:
for which we attempted a best fit as:
$$\k(\t)=-0.01(\pm0.02)-4.8(\pm0.9)\fra1\t\Eq(4.11)$$
with a goodness of $4.03\cdot10^{-4}$. The data are not many because the
experiment is hard (in terms of CPU time).
 
Finally the main quantity of interest, \ie the graph of
$x(p)=\fra1{\t t_0\media{\s}_+}\log\fra{\p_\t(p)}{\p_\t(-p)}$ versus
$p$:

%%%%%%%%%%%%%%%%%%%%%%%%%%%%%%%%%%%%
 
%\input figpr10_ps.tex
\dimen2=25pt \advance\dimen2 by 95pt
\dimen3=103pt \advance\dimen3 by 40pt
\dimenfor{\dimen3}{125pt}
\eqfigfor{
\ins{123pt}{0pt}{$\hbox to20pt{\hfill ${\scriptstyle p}$
\hfill}$}
\ins{0pt}{\dimen2}{$\hbox to25pt{\hfill ${\scriptstyle x(p)}$
\hfill}$}
\ins{71pt}{113pt}{$\hbox to51pt{\hfill ${\scriptstyle \tau=20}$
\hfill}$}
\ins{13pt}{3pt}{$\hbox to 27pt{\hfill${\scriptstyle 0}$
\hfill}$}
\ins{40pt}{3pt}{$\hbox to 27pt{\hfill${\scriptstyle 1}$
\hfill}$}
\ins{68pt}{3pt}{$\hbox to 27pt{\hfill${\scriptstyle 2}$
\hfill}$}
\ins{95pt}{3pt}{$\hbox to 27pt{\hfill${\scriptstyle 3}$
\hfill}$}
\ins{0pt}{20pt}{$\hbox to 20pt{\hfill${\scriptstyle 0}$
\hfill}$}
\ins{0pt}{39pt}{$\hbox to 20pt{\hfill${\scriptstyle 1}$
\hfill}$}
\ins{0pt}{57pt}{$\hbox to 20pt{\hfill${\scriptstyle 2}$
\hfill}$}
\ins{0pt}{75pt}{$\hbox to 20pt{\hfill${\scriptstyle 3}$
\hfill}$}
\ins{0pt}{93pt}{$\hbox to 20pt{\hfill${\scriptstyle 4}$
\hfill}$}
}{figpr100}{\ins{123pt}{0pt}{$\hbox to20pt{\hfill ${\scriptstyle p}$
\hfill}$}
\ins{0pt}{\dimen2}{$\hbox to25pt{\hfill ${\scriptstyle x(p)}$
\hfill}$}
\ins{71pt}{113pt}{$\hbox to51pt{\hfill ${\scriptstyle \tau=40}$
\hfill}$}
\ins{15pt}{3pt}{$\hbox to 21pt{\hfill${\scriptstyle 0.0}$
\hfill}$}
\ins{37pt}{3pt}{$\hbox to 21pt{\hfill${\scriptstyle 0.5}$
\hfill}$}
\ins{58pt}{3pt}{$\hbox to 21pt{\hfill${\scriptstyle 1.0}$
\hfill}$}
\ins{80pt}{3pt}{$\hbox to 21pt{\hfill${\scriptstyle 1.5}$
\hfill}$}
\ins{101pt}{3pt}{$\hbox to 21pt{\hfill${\scriptstyle 2.0}$
\hfill}$}
\ins{0pt}{20pt}{$\hbox to 20pt{\hfill${\scriptstyle 0}$
\hfill}$}
\ins{0pt}{55pt}{$\hbox to 20pt{\hfill${\scriptstyle 1}$
\hfill}$}
\ins{0pt}{91pt}{$\hbox to 20pt{\hfill${\scriptstyle 2}$
\hfill}$}
}{figpr101}{\ins{123pt}{0pt}{$\hbox to20pt{\hfill ${\scriptstyle p}$
\hfill}$}
\ins{0pt}{\dimen2}{$\hbox to25pt{\hfill ${\scriptstyle x(p)}$
\hfill}$}
\ins{71pt}{113pt}{$\hbox to51pt{\hfill ${\scriptstyle \tau=60}$
\hfill}$}
\ins{12pt}{3pt}{$\hbox to 27pt{\hfill${\scriptstyle 0.0}$
\hfill}$}
\ins{39pt}{3pt}{$\hbox to 27pt{\hfill${\scriptstyle 0.5}$
\hfill}$}
\ins{66pt}{3pt}{$\hbox to 27pt{\hfill${\scriptstyle 1.0}$
\hfill}$}
\ins{94pt}{3pt}{$\hbox to 27pt{\hfill${\scriptstyle 1.5}$
\hfill}$}
\ins{0pt}{19pt}{$\hbox to 20pt{\hfill${\scriptstyle 0.0}$
\hfill}$}
\ins{0pt}{43pt}{$\hbox to 20pt{\hfill${\scriptstyle 0.5}$
\hfill}$}
\ins{0pt}{66pt}{$\hbox to 20pt{\hfill${\scriptstyle 1.0}$
\hfill}$}
\ins{0pt}{90pt}{$\hbox to 20pt{\hfill${\scriptstyle 1.5}$
\hfill}$}
}{figpr102}{\ins{123pt}{0pt}{$\hbox to20pt{\hfill ${\scriptstyle p}$
\hfill}$}
\ins{0pt}{\dimen2}{$\hbox to25pt{\hfill ${\scriptstyle x(p)}$
\hfill}$}
\ins{71pt}{113pt}{$\hbox to51pt{\hfill ${\scriptstyle \tau=100}$
\hfill}$}
\ins{1pt}{3pt}{$\hbox to 45pt{\hfill${\scriptstyle 0.0}$
\hfill}$}
\ins{46pt}{3pt}{$\hbox to 45pt{\hfill${\scriptstyle 0.5}$
\hfill}$}
\ins{92pt}{3pt}{$\hbox to 45pt{\hfill${\scriptstyle 1.0}$
\hfill}$}
\ins{0pt}{18pt}{$\hbox to 20pt{\hfill${\scriptstyle 0.0}$
\hfill}$}
\ins{0pt}{53pt}{$\hbox to 20pt{\hfill${\scriptstyle 0.5}$
\hfill}$}
\ins{0pt}{89pt}{$\hbox to 20pt{\hfill${\scriptstyle 1.0}$
\hfill}$}
}{figpr103}{\eq(4.12)}
\*
\didascalia{Fig.7: The graphs for the fluctuation theorem test for
$\t=20,40,60,100$. The dashed straight line is the theoretical
prediction of the fluctuation theorem. The arrows mark the point at
distance $\sqrt{\media{(p-1)^2}}$ from $1$. The error bars are inherited
from the histogram data in Fig.5.}
 
The abscissae axis is discretized and the number of events in which $p$
falls in a given interval is taken as proportional to $\p_\t(p)$.
Therefore Fig.7 is a histogram.
 
It is remarkable that the finite size effects, \ie the manifestation of
important deviations from the linear law for ``small'' $\t$ (as we
interpret them), are here very clear: the case $\t=20$ does not follow
the scaling, in contrast to $\t=40,\,60,\,80,\,100$, ($\t=80$ is not
shown).  \*
 
\1
\0{\bf\S4.1.4 Conclusions, $N=10$:} The case $N=10$ is very similar to
the case $N=2$ as the theory predicts.  From the above data we cannot
exclude that the distribution is Gaussian in the observed range of
values of $p$.  {\it Assuming} that it is Gaussian then the theory gives a
standard deviation $m_2(\t)=\fra2{\t t_0 \media{\s}_+}\=\fra{A}\t$, see
\equ(2.11) and the above experimental data give:
 
$$\eqalign{
A=&25.82\pm.02\quad({\rm Gaussian\ assumption})\cr
A=&25.24\pm.4\quad({\rm experiment})\cr}\Eq(4.13)$$
where the theoretical value is computed from the experimentally measured
$\media{\s}_+, t_0$, while the experiment value is computed from the graphs
for $m_2(\t)$. The error analysis is carried along the same lines as
that of the case $N=2$ above.
 
The standard deviations of $p$ from the value $1$ are for
$\t=20,\,40,\,60,\,80,\,100$ respectively $\sqrt{\fra{A}\t}=1.05$,
$0.79$, $0.66$, $0.57$ and
$0.51$.  Such values, multiplied by $3$, can
be arbitrarily assumed to be the boundary between the small and the
large deviations.
 
The same comments to the case $N=2$ can be made here and therefore the
data, in our opinion, yield a good test of the chaotic assumption in the
case $N=10$ too.
 
\* \0{\bf\S4.3.1 Semi-periodic boundary conditions, $N=2$}: The results
are very similar in the case of semi-periodic boundary conditions and
the following two graphs give the experimental values over $10^8$
collisions (with the walls, other particles or obstacles) for the
probability distribution $\p_\t(p)$ with $\t=20,100$; the $\p_\t(p)$ at
intermediate values of $\t$ have also been measured ($\t=40,60,80$) but
we do not give the graphs.
 
\vskip .5cm
 
%%%%%%%%%%%%%%%%%%%%%%%%%%%%

%\input figbh2_ps.tex
\dimen2=25pt \advance\dimen2 by 95pt
\dimen3=103pt \advance\dimen3 by 40pt
\eqfigbis{\dimen3}{125pt}{
\ins{123pt}{0pt}{$\hbox to20pt{\hfill ${\scriptstyle p}$
\hfill}$}
\ins{0pt}{\dimen2}{$\hbox to25pt{\hfill ${\scriptstyle \pi(p)}$
\hfill}$}
\ins{71pt}{113pt}{$\hbox to51pt{\hfill ${\scriptstyle \tau=20}$
\hfill}$}
\ins{15pt}{3pt}{$\hbox to 19pt{\hfill${\scriptstyle -6}$
\hfill}$}
\ins{34pt}{3pt}{$\hbox to 19pt{\hfill${\scriptstyle -3}$
\hfill}$}
\ins{54pt}{3pt}{$\hbox to 19pt{\hfill${\scriptstyle 0}$
\hfill}$}
\ins{73pt}{3pt}{$\hbox to 19pt{\hfill${\scriptstyle 3}$
\hfill}$}
\ins{92pt}{3pt}{$\hbox to 19pt{\hfill${\scriptstyle 6}$
\hfill}$}
\ins{112pt}{3pt}{$\hbox to 19pt{\hfill${\scriptstyle 9}$
\hfill}$}
\ins{0pt}{21pt}{$\hbox to 20pt{\hfill${\scriptstyle 0.0}$
\hfill}$}
\ins{0pt}{35pt}{$\hbox to 20pt{\hfill${\scriptstyle 0.1}$
\hfill}$}
\ins{0pt}{48pt}{$\hbox to 20pt{\hfill${\scriptstyle 0.2}$
\hfill}$}
\ins{0pt}{61pt}{$\hbox to 20pt{\hfill${\scriptstyle 0.3}$
\hfill}$}
\ins{0pt}{74pt}{$\hbox to 20pt{\hfill${\scriptstyle 0.4}$
\hfill}$}
\ins{0pt}{87pt}{$\hbox to 20pt{\hfill${\scriptstyle 0.5}$
\hfill}$}
\ins{0pt}{100pt}{$\hbox to 20pt{\hfill${\scriptstyle 0.6}$
\hfill}$}
}{figbh20}{\ins{123pt}{0pt}{$\hbox to20pt{\hfill ${\scriptstyle p}$
\hfill}$}
\ins{0pt}{\dimen2}{$\hbox to25pt{\hfill ${\scriptstyle \pi(p)}$
\hfill}$}
\ins{71pt}{113pt}{$\hbox to51pt{\hfill ${\scriptstyle \tau=100}$
\hfill}$}
\ins{15pt}{3pt}{$\hbox to 19pt{\hfill${\scriptstyle -6}$
\hfill}$}
\ins{34pt}{3pt}{$\hbox to 19pt{\hfill${\scriptstyle -3}$
\hfill}$}
\ins{54pt}{3pt}{$\hbox to 19pt{\hfill${\scriptstyle 0}$
\hfill}$}
\ins{73pt}{3pt}{$\hbox to 19pt{\hfill${\scriptstyle 3}$
\hfill}$}
\ins{92pt}{3pt}{$\hbox to 19pt{\hfill${\scriptstyle 6}$
\hfill}$}
\ins{112pt}{3pt}{$\hbox to 19pt{\hfill${\scriptstyle 9}$
\hfill}$}
\ins{0pt}{21pt}{$\hbox to 20pt{\hfill${\scriptstyle 0.0}$
\hfill}$}
\ins{0pt}{35pt}{$\hbox to 20pt{\hfill${\scriptstyle 0.1}$
\hfill}$}
\ins{0pt}{48pt}{$\hbox to 20pt{\hfill${\scriptstyle 0.2}$
\hfill}$}
\ins{0pt}{61pt}{$\hbox to 20pt{\hfill${\scriptstyle 0.3}$
\hfill}$}
\ins{0pt}{74pt}{$\hbox to 20pt{\hfill${\scriptstyle 0.4}$
\hfill}$}
\ins{0pt}{87pt}{$\hbox to 20pt{\hfill${\scriptstyle 0.5}$
\hfill}$}
\ins{0pt}{100pt}{$\hbox to 20pt{\hfill${\scriptstyle 0.6}$
\hfill}$}
}{figbh21}{\eq(4.14)}
 
%%%%%%%%%%%%%%%%%%%%%%%%%%%
 
\didascalia{Fig.8: The histograms for $\p_\t(p)$ for
$\t=20,100$.}
 
The errors are treated, naturally, as in the corresponding previous
cases.
 
The next two figures give the graphs of the $x(p)$ in \equ(4.5), which
is expected to be the graph of $x(p)=p$. Again we give only the two
extreme cases, $\t=20,100$ we have been able to measure: the agreement
with the ``theory'' is as excellent in the intermediate cases
($\t=40,60,80$). The errors are slightly smaller than in the
previous experiments because of the difference in the length of the
computed trajectory.
 
\1
\dimen2=25pt \advance\dimen2 by 95pt
\dimen3=103pt \advance\dimen3 by 40pt
\eqfigbis{\dimen3}{125pt}{
\ins{123pt}{0pt}{$\hbox to20pt{\hfill ${\scriptstyle p}$
\hfill}$}
\ins{0pt}{\dimen2}{$\hbox to25pt{\hfill ${\scriptstyle x(p)}$
\hfill}$}
\ins{71pt}{113pt}{$\hbox to51pt{\hfill ${\scriptstyle \tau=20}$
\hfill}$}
\ins{20pt}{3pt}{$\hbox to 15pt{\hfill${\scriptstyle 0}$
\hfill}$}
\ins{35pt}{3pt}{$\hbox to 15pt{\hfill${\scriptstyle 1}$
\hfill}$}
\ins{51pt}{3pt}{$\hbox to 15pt{\hfill${\scriptstyle 2}$
\hfill}$}
\ins{66pt}{3pt}{$\hbox to 15pt{\hfill${\scriptstyle 3}$
\hfill}$}
\ins{81pt}{3pt}{$\hbox to 15pt{\hfill${\scriptstyle 4}$
\hfill}$}
\ins{97pt}{3pt}{$\hbox to 15pt{\hfill${\scriptstyle 5}$
\hfill}$}
\ins{112pt}{3pt}{$\hbox to 15pt{\hfill${\scriptstyle 6}$
\hfill}$}
\ins{0pt}{21pt}{$\hbox to 20pt{\hfill${\scriptstyle 0}$
\hfill}$}
\ins{0pt}{31pt}{$\hbox to 20pt{\hfill${\scriptstyle 1}$
\hfill}$}
\ins{0pt}{41pt}{$\hbox to 20pt{\hfill${\scriptstyle 2}$
\hfill}$}
\ins{0pt}{50pt}{$\hbox to 20pt{\hfill${\scriptstyle 3}$
\hfill}$}
\ins{0pt}{60pt}{$\hbox to 20pt{\hfill${\scriptstyle 4}$
\hfill}$}
\ins{0pt}{70pt}{$\hbox to 20pt{\hfill${\scriptstyle 5}$
\hfill}$}
\ins{0pt}{80pt}{$\hbox to 20pt{\hfill${\scriptstyle 6}$
\hfill}$}
\ins{0pt}{89pt}{$\hbox to 20pt{\hfill${\scriptstyle 7}$
\hfill}$}
\ins{0pt}{99pt}{$\hbox to 20pt{\hfill${\scriptstyle 8}$
\hfill}$}
}{figbr20}{\ins{123pt}{0pt}{$\hbox to20pt{\hfill ${\scriptstyle p}$
\hfill}$}
\ins{0pt}{\dimen2}{$\hbox to25pt{\hfill ${\scriptstyle x(p)}$
\hfill}$}
\ins{71pt}{113pt}{$\hbox to51pt{\hfill ${\scriptstyle \tau=100}$
\hfill}$}
\ins{13pt}{3pt}{$\hbox to 25pt{\hfill${\scriptstyle 0.0}$
\hfill}$}
\ins{38pt}{3pt}{$\hbox to 25pt{\hfill${\scriptstyle 0.5}$
\hfill}$}
\ins{64pt}{3pt}{$\hbox to 25pt{\hfill${\scriptstyle 1.0}$
\hfill}$}
\ins{89pt}{3pt}{$\hbox to 25pt{\hfill${\scriptstyle 1.5}$
\hfill}$}
\ins{0pt}{20pt}{$\hbox to 20pt{\hfill${\scriptstyle 0}$
\hfill}$}
\ins{0pt}{53pt}{$\hbox to 20pt{\hfill${\scriptstyle 1}$
\hfill}$}
\ins{0pt}{85pt}{$\hbox to 20pt{\hfill${\scriptstyle 2}$
\hfill}$}
}{figbr21}{\eq(4.15)}
 
%%%%%%%%%%%%%%%%%%%%%%%%%%%%%%%
\*
\didascalia{Fig.9: The graphs for the fluctuation theorem test, case of
$\fra12(pbc),\,N=2$; the dashed line is the fluctuation theorem
prediction, $\t=20,100$. The arrows mark the point at
distance $\sqrt{\media{(p-1)^2}}$ from $1$.}
 
We have also measured the kurtosis and the decay of the second and third
moments: the results are very similar to the ones in the periodic
boundary conditions case and can be commented in the same way.
 
\vskip .5cm
%%%%%%%%%%%%%%%%%%%%%%%%%%%%%%%%%%
 
%\input figbv2_ps.tex
\dimen2=25pt \advance\dimen2 by 95pt
\dimen3=103pt \advance\dimen3 by 40pt
\dimenfor{\dimen3}{125pt}
\eqfigter{
\ins{123pt}{0pt}{$\hbox to20pt{\hfill ${\scriptstyle \tau^{-1}}$
\hfill}$}
\ins{0pt}{\dimen2}{$\hbox to25pt{\hfill ${\scriptstyle m_2(\tau)}$
\hfill}$}
\ins{14pt}{3pt}{$\hbox to 26pt{\hfill${\scriptstyle 0.00}$
\hfill}$}
\ins{40pt}{3pt}{$\hbox to 26pt{\hfill${\scriptstyle 0.03}$
\hfill}$}
\ins{67pt}{3pt}{$\hbox to 26pt{\hfill${\scriptstyle 0.06}$
\hfill}$}
\ins{93pt}{3pt}{$\hbox to 26pt{\hfill${\scriptstyle 0.09}$
\hfill}$}
\ins{0pt}{21pt}{$\hbox to 20pt{\hfill${\scriptstyle 0}$
\hfill}$}
\ins{0pt}{38pt}{$\hbox to 20pt{\hfill${\scriptstyle 1}$
\hfill}$}
\ins{0pt}{56pt}{$\hbox to 20pt{\hfill${\scriptstyle 2}$
\hfill}$}
\ins{0pt}{74pt}{$\hbox to 20pt{\hfill${\scriptstyle 3}$
\hfill}$}
\ins{0pt}{91pt}{$\hbox to 20pt{\hfill${\scriptstyle 4}$
\hfill}$}
}{figbv20}{\ins{123pt}{0pt}{$\hbox to20pt{\hfill ${\scriptstyle \tau^{-2}}$
\hfill}$}
\ins{0pt}{\dimen2}{$\hbox to25pt{\hfill ${\scriptstyle m_3(\tau)}$
\hfill}$}
\ins{7pt}{3pt}{$\hbox to 43pt{\hfill${\scriptstyle 0.000}$
\hfill}$}
\ins{50pt}{3pt}{$\hbox to 43pt{\hfill${\scriptstyle 0.005}$
\hfill}$}
\ins{93pt}{3pt}{$\hbox to 43pt{\hfill${\scriptstyle 0.010}$
\hfill}$}
\ins{0pt}{21pt}{$\hbox to 20pt{\hfill${\scriptstyle 0.0}$
\hfill}$}
\ins{0pt}{34pt}{$\hbox to 20pt{\hfill${\scriptstyle 0.2}$
\hfill}$}
\ins{0pt}{46pt}{$\hbox to 20pt{\hfill${\scriptstyle 0.4}$
\hfill}$}
\ins{0pt}{59pt}{$\hbox to 20pt{\hfill${\scriptstyle 0.6}$
\hfill}$}
\ins{0pt}{71pt}{$\hbox to 20pt{\hfill${\scriptstyle 0.8}$
\hfill}$}
\ins{0pt}{83pt}{$\hbox to 20pt{\hfill${\scriptstyle 1.0}$
\hfill}$}
\ins{0pt}{96pt}{$\hbox to 20pt{\hfill${\scriptstyle 1.2}$
\hfill}$}
\ins{0pt}{108pt}{$\hbox to 20pt{\hfill${\scriptstyle 1.4}$
\hfill}$}
}{figbv21}{\ins{123pt}{0pt}{$\hbox to20pt{\hfill ${\scriptstyle \tau^{-1}}$
\hfill}$}
\ins{0pt}{\dimen2}{$\hbox to25pt{\hfill ${\scriptstyle \k(\tau)}$
\hfill}$}
\ins{14pt}{3pt}{$\hbox to 26pt{\hfill${\scriptstyle 0.00}$
\hfill}$}
\ins{40pt}{3pt}{$\hbox to 26pt{\hfill${\scriptstyle 0.03}$
\hfill}$}
\ins{67pt}{3pt}{$\hbox to 26pt{\hfill${\scriptstyle 0.06}$
\hfill}$}
\ins{93pt}{3pt}{$\hbox to 26pt{\hfill${\scriptstyle 0.09}$
\hfill}$}
\ins{0pt}{22pt}{$\hbox to 20pt{\hfill${\scriptstyle -0.1}$
\hfill}$}
\ins{0pt}{42pt}{$\hbox to 20pt{\hfill${\scriptstyle 0.0}$
\hfill}$}
\ins{0pt}{61pt}{$\hbox to 20pt{\hfill${\scriptstyle 0.1}$
\hfill}$}
\ins{0pt}{80pt}{$\hbox to 20pt{\hfill${\scriptstyle 0.2}$
\hfill}$}
\ins{0pt}{99pt}{$\hbox to 20pt{\hfill${\scriptstyle 0.3}$
\hfill}$\hfil}
}{figbv22}{\eq(4.16)}
\*
\didascalia{Fig.10: The second moment, the third moment and the
kurtosis.}
 
The fits are:
$$\eqalign{
&\txt m_2(\t)=-0.00(\pm 0.002)+45.1(\pm 0.01){1\over \t}\cr
&\txt m_3(\t)=0.001(\pm 0.004)+122.6(\pm 10.4){1\over \t^2}\cr
&\txt \k(\t)=-0.009(\pm 0.04)+2.7(\pm 0.8){1\over \t}\cr
}\Eq(4.17)
$$
\0and the goodness (see above, and the Appendix, for the definition) are
$3\cdot 10^{-3}$, $2\cdot 10^{-3}$ and $1\cdot 10^{-2}$, respectively.
\1

As in the previous cases we can assume that is a Gaussian and we see that
the results are:
 
$$\eqalign{
&A=43.8\pm 0.03 \quad{\rm Gaussian\ assumption}\cr
&A=45.1\pm 0.01 \quad{\rm experiment}\cr
}\Eq(4.18)$$
As in the case of periodic boundary conditions the distribution is quite
close to a Gaussian, although it cannot be such, strictly speaking (for
the reasons already discussed).
\*
 
\0{\it\S4.3.4 Semi-periodic boundary conditions and $10$ particles:}
We have also considered a further $N=10$ particles system, but at a
density quite different from the previous one.
 
As in the previous case we need to take $\t$ rather large to see the
fluctuation law $x(p)=p$, see \equ(4.5), to hold. But it turns out that
although at $\t=20$ ``finite size effects'' are still visible and very
strong, already at $\t=40$ they become essentially not visible.
 
Therefore we give the histogram for $\p_\t(p)$ at $\t=20,40,60,100$:
 
\*
\dimen2=25pt \advance\dimen2 by 95pt
\dimen3=103pt \advance\dimen3 by 40pt
\dimenfor{\dimen3}{125pt}
\eqfigfor{
\ins{123pt}{0pt}{$\hbox to20pt{\hfill ${\scriptstyle p}$
\hfill}$}
\ins{0pt}{\dimen2}{$\hbox to25pt{\hfill ${\scriptstyle \pi(p)}$
\hfill}$}
\ins{71pt}{113pt}{$\hbox to51pt{\hfill ${\scriptstyle \tau=20}$
\hfill}$}
\ins{21pt}{3pt}{$\hbox to 15pt{\hfill${\scriptstyle -4}$
\hfill}$}
\ins{37pt}{3pt}{$\hbox to 15pt{\hfill${\scriptstyle -2}$
\hfill}$}
\ins{53pt}{3pt}{$\hbox to 15pt{\hfill${\scriptstyle 0}$
\hfill}$}
\ins{69pt}{3pt}{$\hbox to 15pt{\hfill${\scriptstyle 2}$
\hfill}$}
\ins{85pt}{3pt}{$\hbox to 15pt{\hfill${\scriptstyle 4}$
\hfill}$}
\ins{101pt}{3pt}{$\hbox to 15pt{\hfill${\scriptstyle 6}$
\hfill}$}
\ins{0pt}{21pt}{$\hbox to 20pt{\hfill${\scriptstyle 0.0}$
\hfill}$}
\ins{0pt}{32pt}{$\hbox to 20pt{\hfill${\scriptstyle 0.1}$
\hfill}$}
\ins{0pt}{43pt}{$\hbox to 20pt{\hfill${\scriptstyle 0.2}$
\hfill}$}
\ins{0pt}{54pt}{$\hbox to 20pt{\hfill${\scriptstyle 0.3}$
\hfill}$}
\ins{0pt}{65pt}{$\hbox to 20pt{\hfill${\scriptstyle 0.4}$
\hfill}$}
\ins{0pt}{76pt}{$\hbox to 20pt{\hfill${\scriptstyle 0.5}$
\hfill}$}
\ins{0pt}{87pt}{$\hbox to 20pt{\hfill${\scriptstyle 0.6}$
\hfill}$}
\ins{0pt}{98pt}{$\hbox to 20pt{\hfill${\scriptstyle 0.7}$
\hfill}$}
}{figbh100}{\ins{123pt}{0pt}{$\hbox to20pt{\hfill ${\scriptstyle p}$
\hfill}$}
\ins{0pt}{\dimen2}{$\hbox to25pt{\hfill ${\scriptstyle \pi(p)}$
\hfill}$}
\ins{71pt}{113pt}{$\hbox to51pt{\hfill ${\scriptstyle \tau=40}$
\hfill}$}
\ins{21pt}{3pt}{$\hbox to 15pt{\hfill${\scriptstyle -4}$
\hfill}$}
\ins{37pt}{3pt}{$\hbox to 15pt{\hfill${\scriptstyle -2}$
\hfill}$}
\ins{53pt}{3pt}{$\hbox to 15pt{\hfill${\scriptstyle 0}$
\hfill}$}
\ins{69pt}{3pt}{$\hbox to 15pt{\hfill${\scriptstyle 2}$
\hfill}$}
\ins{85pt}{3pt}{$\hbox to 15pt{\hfill${\scriptstyle 4}$
\hfill}$}
\ins{101pt}{3pt}{$\hbox to 15pt{\hfill${\scriptstyle 6}$
\hfill}$}
\ins{0pt}{21pt}{$\hbox to 20pt{\hfill${\scriptstyle 0.0}$
\hfill}$}
\ins{0pt}{32pt}{$\hbox to 20pt{\hfill${\scriptstyle 0.1}$
\hfill}$}
\ins{0pt}{43pt}{$\hbox to 20pt{\hfill${\scriptstyle 0.2}$
\hfill}$}
\ins{0pt}{54pt}{$\hbox to 20pt{\hfill${\scriptstyle 0.3}$
\hfill}$}
\ins{0pt}{65pt}{$\hbox to 20pt{\hfill${\scriptstyle 0.4}$
\hfill}$}
\ins{0pt}{76pt}{$\hbox to 20pt{\hfill${\scriptstyle 0.5}$
\hfill}$}
\ins{0pt}{87pt}{$\hbox to 20pt{\hfill${\scriptstyle 0.6}$
\hfill}$}
\ins{0pt}{98pt}{$\hbox to 20pt{\hfill${\scriptstyle 0.7}$
\hfill}$}
}{figbh101}{\ins{123pt}{0pt}{$\hbox to20pt{\hfill ${\scriptstyle p}$
\hfill}$}
\ins{0pt}{\dimen2}{$\hbox to25pt{\hfill ${\scriptstyle \pi(p)}$
\hfill}$}
\ins{71pt}{113pt}{$\hbox to51pt{\hfill ${\scriptstyle \tau=60}$
\hfill}$}
\ins{21pt}{3pt}{$\hbox to 15pt{\hfill${\scriptstyle -4}$
\hfill}$}
\ins{37pt}{3pt}{$\hbox to 15pt{\hfill${\scriptstyle -2}$
\hfill}$}
\ins{53pt}{3pt}{$\hbox to 15pt{\hfill${\scriptstyle 0}$
\hfill}$}
\ins{69pt}{3pt}{$\hbox to 15pt{\hfill${\scriptstyle 2}$
\hfill}$}
\ins{85pt}{3pt}{$\hbox to 15pt{\hfill${\scriptstyle 4}$
\hfill}$}
\ins{101pt}{3pt}{$\hbox to 15pt{\hfill${\scriptstyle 6}$
\hfill}$}
\ins{0pt}{21pt}{$\hbox to 20pt{\hfill${\scriptstyle 0.0}$
\hfill}$}
\ins{0pt}{32pt}{$\hbox to 20pt{\hfill${\scriptstyle 0.1}$
\hfill}$}
\ins{0pt}{43pt}{$\hbox to 20pt{\hfill${\scriptstyle 0.2}$
\hfill}$}
\ins{0pt}{54pt}{$\hbox to 20pt{\hfill${\scriptstyle 0.3}$
\hfill}$}
\ins{0pt}{65pt}{$\hbox to 20pt{\hfill${\scriptstyle 0.4}$
\hfill}$}
\ins{0pt}{76pt}{$\hbox to 20pt{\hfill${\scriptstyle 0.5}$
\hfill}$}
\ins{0pt}{87pt}{$\hbox to 20pt{\hfill${\scriptstyle 0.6}$
\hfill}$}
\ins{0pt}{98pt}{$\hbox to 20pt{\hfill${\scriptstyle 0.7}$
\hfill}$}
}{figbh102}{\ins{123pt}{0pt}{$\hbox to20pt{\hfill ${\scriptstyle p}$
\hfill}$}
\ins{0pt}{\dimen2}{$\hbox to25pt{\hfill ${\scriptstyle \pi(p)}$
\hfill}$}
\ins{71pt}{113pt}{$\hbox to51pt{\hfill ${\scriptstyle \tau=100}$
\hfill}$}
\ins{21pt}{3pt}{$\hbox to 15pt{\hfill${\scriptstyle -4}$
\hfill}$}
\ins{37pt}{3pt}{$\hbox to 15pt{\hfill${\scriptstyle -2}$
\hfill}$}
\ins{53pt}{3pt}{$\hbox to 15pt{\hfill${\scriptstyle 0}$
\hfill}$}
\ins{69pt}{3pt}{$\hbox to 15pt{\hfill${\scriptstyle 2}$
\hfill}$}
\ins{85pt}{3pt}{$\hbox to 15pt{\hfill${\scriptstyle 4}$
\hfill}$}
\ins{101pt}{3pt}{$\hbox to 15pt{\hfill${\scriptstyle 6}$
\hfill}$}
\ins{0pt}{21pt}{$\hbox to 20pt{\hfill${\scriptstyle 0.0}$
\hfill}$}
\ins{0pt}{32pt}{$\hbox to 20pt{\hfill${\scriptstyle 0.1}$
\hfill}$}
\ins{0pt}{43pt}{$\hbox to 20pt{\hfill${\scriptstyle 0.2}$
\hfill}$}
\ins{0pt}{54pt}{$\hbox to 20pt{\hfill${\scriptstyle 0.3}$
\hfill}$}
\ins{0pt}{65pt}{$\hbox to 20pt{\hfill${\scriptstyle 0.4}$
\hfill}$}
\ins{0pt}{76pt}{$\hbox to 20pt{\hfill${\scriptstyle 0.5}$
\hfill}$}
\ins{0pt}{87pt}{$\hbox to 20pt{\hfill${\scriptstyle 0.6}$
\hfill}$}
\ins{0pt}{98pt}{$\hbox to 20pt{\hfill${\scriptstyle 0.7}$
\hfill}$}
}{figbh103}{\eq(4.19)}
\*
\didascalia{\it Fig.11: Histograms for $\p_\t(p)$ at $\t=20$
and $\t=100$.}

\1 
\0and the graphs of $x(p)$ at $\t=20,40,80,100$:
 
\vskip .5cm
 
%%%%%%%%%%%%%%%%%%%%%%%

%\input figbr10_ps.tex
\dimen2=25pt \advance\dimen2 by 95pt
\dimen3=103pt \advance\dimen3 by 40pt
\dimenfor{\dimen3}{125pt}
\eqfigfor{
\ins{123pt}{0pt}{$\hbox to20pt{\hfill ${\scriptstyle p}$
\hfill}$}
\ins{0pt}{\dimen2}{$\hbox to25pt{\hfill ${\scriptstyle x(p)}$
\hfill}$}
\ins{71pt}{113pt}{$\hbox to51pt{\hfill ${\scriptstyle \tau=20}$
\hfill}$}
\ins{18pt}{3pt}{$\hbox to 19pt{\hfill${\scriptstyle 0}$
\hfill}$}
\ins{37pt}{3pt}{$\hbox to 19pt{\hfill${\scriptstyle 1}$
\hfill}$}
\ins{56pt}{3pt}{$\hbox to 19pt{\hfill${\scriptstyle 2}$
\hfill}$}
\ins{75pt}{3pt}{$\hbox to 19pt{\hfill${\scriptstyle 3}$
\hfill}$}
\ins{94pt}{3pt}{$\hbox to 19pt{\hfill${\scriptstyle 4}$
\hfill}$}
\ins{0pt}{21pt}{$\hbox to 20pt{\hfill${\scriptstyle 0}$
\hfill}$}
\ins{0pt}{34pt}{$\hbox to 20pt{\hfill${\scriptstyle 1}$
\hfill}$}
\ins{0pt}{48pt}{$\hbox to 20pt{\hfill${\scriptstyle 2}$
\hfill}$}
\ins{0pt}{61pt}{$\hbox to 20pt{\hfill${\scriptstyle 3}$
\hfill}$}
\ins{0pt}{75pt}{$\hbox to 20pt{\hfill${\scriptstyle 4}$
\hfill}$}
\ins{0pt}{88pt}{$\hbox to 20pt{\hfill${\scriptstyle 5}$
\hfill}$}
\ins{0pt}{102pt}{$\hbox to 20pt{\hfill${\scriptstyle 6}$
\hfill}$}
}{figbr100}{\ins{123pt}{0pt}{$\hbox to20pt{\hfill ${\scriptstyle p}$
\hfill}$}
\ins{0pt}{\dimen2}{$\hbox to25pt{\hfill ${\scriptstyle x(p)}$
\hfill}$}
\ins{71pt}{113pt}{$\hbox to51pt{\hfill ${\scriptstyle \tau=40}$
\hfill}$}
\ins{19pt}{3pt}{$\hbox to 15pt{\hfill${\scriptstyle 0.0}$
\hfill}$}
\ins{35pt}{3pt}{$\hbox to 15pt{\hfill${\scriptstyle 0.5}$
\hfill}$}
\ins{51pt}{3pt}{$\hbox to 15pt{\hfill${\scriptstyle 1.0}$
\hfill}$}
\ins{67pt}{3pt}{$\hbox to 15pt{\hfill${\scriptstyle 1.5}$
\hfill}$}
\ins{83pt}{3pt}{$\hbox to 15pt{\hfill${\scriptstyle 2.0}$
\hfill}$}
\ins{99pt}{3pt}{$\hbox to 15pt{\hfill${\scriptstyle 2.5}$
\hfill}$}
\ins{115pt}{3pt}{$\hbox to 15pt{\hfill${\scriptstyle 3.0}$
\hfill}$}
\ins{0pt}{20pt}{$\hbox to 20pt{\hfill${\scriptstyle 0}$
\hfill}$}
\ins{0pt}{44pt}{$\hbox to 20pt{\hfill${\scriptstyle 1}$
\hfill}$}
\ins{0pt}{68pt}{$\hbox to 20pt{\hfill${\scriptstyle 2}$
\hfill}$}
\ins{0pt}{93pt}{$\hbox to 20pt{\hfill${\scriptstyle 3}$
\hfill}$}
}{figbr101}{\ins{123pt}{0pt}{$\hbox to20pt{\hfill ${\scriptstyle p}$
\hfill}$}
\ins{0pt}{\dimen2}{$\hbox to25pt{\hfill ${\scriptstyle x(p)}$
\hfill}$}
\ins{71pt}{113pt}{$\hbox to51pt{\hfill ${\scriptstyle \tau=60}$
\hfill}$}
\ins{15pt}{3pt}{$\hbox to 21pt{\hfill${\scriptstyle 0.0}$
\hfill}$}
\ins{37pt}{3pt}{$\hbox to 21pt{\hfill${\scriptstyle 0.5}$
\hfill}$}
\ins{58pt}{3pt}{$\hbox to 21pt{\hfill${\scriptstyle 1.0}$
\hfill}$}
\ins{80pt}{3pt}{$\hbox to 21pt{\hfill${\scriptstyle 1.5}$
\hfill}$}
\ins{102pt}{3pt}{$\hbox to 21pt{\hfill${\scriptstyle 2.0}$
\hfill}$}
\ins{0pt}{20pt}{$\hbox to 20pt{\hfill${\scriptstyle 0.0}$
\hfill}$}
\ins{0pt}{35pt}{$\hbox to 20pt{\hfill${\scriptstyle 0.5}$
\hfill}$}
\ins{0pt}{50pt}{$\hbox to 20pt{\hfill${\scriptstyle 1.0}$
\hfill}$}
\ins{0pt}{65pt}{$\hbox to 20pt{\hfill${\scriptstyle 1.5}$
\hfill}$}
\ins{0pt}{80pt}{$\hbox to 20pt{\hfill${\scriptstyle 2.0}$
\hfill}$}
\ins{0pt}{94pt}{$\hbox to 20pt{\hfill${\scriptstyle 2.5}$
\hfill}$}
}{figbr102}{\ins{123pt}{0pt}{$\hbox to20pt{\hfill ${\scriptstyle p}$
\hfill}$}
\ins{0pt}{\dimen2}{$\hbox to25pt{\hfill ${\scriptstyle x(p)}$
\hfill}$}
\ins{71pt}{113pt}{$\hbox to51pt{\hfill ${\scriptstyle \tau=100}$
\hfill}$}
\ins{8pt}{3pt}{$\hbox to 33pt{\hfill${\scriptstyle 0.0}$
\hfill}$}
\ins{41pt}{3pt}{$\hbox to 33pt{\hfill${\scriptstyle 0.5}$
\hfill}$}
\ins{75pt}{3pt}{$\hbox to 33pt{\hfill${\scriptstyle 1.0}$
\hfill}$}
\ins{0pt}{19pt}{$\hbox to 20pt{\hfill${\scriptstyle 0.0}$
\hfill}$}
\ins{0pt}{42pt}{$\hbox to 20pt{\hfill${\scriptstyle 0.5}$
\hfill}$}
\ins{0pt}{65pt}{$\hbox to 20pt{\hfill${\scriptstyle 1.0}$
\hfill}$}
\ins{0pt}{89pt}{$\hbox to 20pt{\hfill${\scriptstyle 1.5}$
\hfill}$}
}{figbr103}{\eq(4.20)}
 
%%%%%%%%%%%%%%%%%%%%%%%
 
\*
\didascalia{Fig.12: The linear fluctuation test, $\t=20,40,80,100$. The
dashed line is the fluctuation theorem prediction for $\t=+\io$. The
arrows mark the point at distance $\sqrt{\media{(p-1)^2}}$ from $1$.}
 
\0which illustrate well the fact that there is a visible finite size
effect that very soon becomes not observable.
 
The experiments were carried out over $10^8$ collisions. The
following graphs give the kurtosis and the decay of the second and third
moments.
\1

\*
\dimen2=25pt \advance\dimen2 by 95pt
\dimen3=103pt \advance\dimen3 by 40pt
\dimenfor{\dimen3}{125pt}
\eqfigter{
\ins{123pt}{0pt}{$\hbox to20pt{\hfill ${\scriptstyle \tau^{-1}}$
\hfill}$}
\ins{0pt}{\dimen2}{$\hbox to25pt{\hfill ${\scriptstyle m_2(\tau)}$
\hfill}$}
\ins{14pt}{3pt}{$\hbox to 26pt{\hfill${\scriptstyle 0.00}$
\hfill}$}
\ins{40pt}{3pt}{$\hbox to 26pt{\hfill${\scriptstyle 0.03}$
\hfill}$}
\ins{67pt}{3pt}{$\hbox to 26pt{\hfill${\scriptstyle 0.06}$
\hfill}$}
\ins{93pt}{3pt}{$\hbox to 26pt{\hfill${\scriptstyle 0.09}$
\hfill}$}
\ins{0pt}{20pt}{$\hbox to 20pt{\hfill${\scriptstyle 0.0}$
\hfill}$}
\ins{0pt}{36pt}{$\hbox to 20pt{\hfill${\scriptstyle 0.4}$
\hfill}$}
\ins{0pt}{51pt}{$\hbox to 20pt{\hfill${\scriptstyle 0.8}$
\hfill}$}
\ins{0pt}{67pt}{$\hbox to 20pt{\hfill${\scriptstyle 1.2}$
\hfill}$}
\ins{0pt}{82pt}{$\hbox to 20pt{\hfill${\scriptstyle 1.6}$
\hfill}$}
\ins{0pt}{98pt}{$\hbox to 20pt{\hfill${\scriptstyle 2.0}$
\hfill}$}
}{figbv100}{\ins{123pt}{0pt}{$\hbox to20pt{\hfill ${\scriptstyle \tau^{-2}}$
\hfill}$}
\ins{0pt}{\dimen2}{$\hbox to25pt{\hfill ${\scriptstyle m_3(\tau)}$
\hfill}$}
\ins{7pt}{3pt}{$\hbox to 43pt{\hfill${\scriptstyle 0.000}$
\hfill}$}
\ins{50pt}{3pt}{$\hbox to 43pt{\hfill${\scriptstyle 0.005}$
\hfill}$}
\ins{93pt}{3pt}{$\hbox to 43pt{\hfill${\scriptstyle 0.010}$
\hfill}$}
\ins{0pt}{21pt}{$\hbox to 20pt{\hfill${\scriptstyle 0.0}$
\hfill}$}
\ins{0pt}{38pt}{$\hbox to 20pt{\hfill${\scriptstyle 0.2}$
\hfill}$}
\ins{0pt}{56pt}{$\hbox to 20pt{\hfill${\scriptstyle 0.4}$
\hfill}$}
\ins{0pt}{73pt}{$\hbox to 20pt{\hfill${\scriptstyle 0.6}$
\hfill}$}
\ins{0pt}{90pt}{$\hbox to 20pt{\hfill${\scriptstyle 0.8}$
\hfill}$}
\ins{0pt}{107pt}{$\hbox to 20pt{\hfill${\scriptstyle 1.0}$
\hfill}$}
}{figbv101}{\ins{123pt}{0pt}{$\hbox to20pt{\hfill ${\scriptstyle \tau^{-1}}$
\hfill}$}
\ins{0pt}{\dimen2}{$\hbox to25pt{\hfill ${\scriptstyle \k(\tau)}$
\hfill}$}
\ins{14pt}{3pt}{$\hbox to 26pt{\hfill${\scriptstyle 0.00}$
\hfill}$}
\ins{40pt}{3pt}{$\hbox to 26pt{\hfill${\scriptstyle 0.03}$
\hfill}$}
\ins{67pt}{3pt}{$\hbox to 26pt{\hfill${\scriptstyle 0.06}$
\hfill}$}
\ins{93pt}{3pt}{$\hbox to 26pt{\hfill${\scriptstyle 0.09}$
\hfill}$}
\ins{0pt}{10pt}{$\hbox to 20pt{\hfill${\scriptstyle -0.2}$
\hfill}$}
\ins{0pt}{29pt}{$\hbox to 20pt{\hfill${\scriptstyle 0.0}$
\hfill}$}
\ins{0pt}{48pt}{$\hbox to 20pt{\hfill${\scriptstyle 0.2}$
\hfill}$}
\ins{0pt}{67pt}{$\hbox to 20pt{\hfill${\scriptstyle 0.4}$
\hfill}$}
\ins{0pt}{86pt}{$\hbox to 20pt{\hfill${\scriptstyle 0.6}$
\hfill}$}
\ins{0pt}{106pt}{$\hbox to 20pt{\hfill${\scriptstyle 0.8}$
\hfill}$}
}{figbv102}{\eq(4.21)}
 
%%%%%%%%%%%%%%%%%%%%%%%%%%%%%%
 
\didascalia{\it Fig.13: The decay of the $2^d$ and $3^d$ moments and of
the kurtosis.}
 
The corresponding fits are, excluding the last point that is clearly a
finite size effect:
 
$$\eqalign{
&m_2(\t)=-0.002(\pm 0.0005)+30.7(\pm 0.1){1\over \t}\cr
&m_3(\t)=0.0001(\pm 0.001)+84.3(\pm 11.3){1\over \t^2}\cr
&\k(\t)=-0.00(\pm 0.07)+2.3(\pm 5.7){1\over \t}\cr}\Eq(4.22)
$$
and the goodness are $2\cdot 10^{-3}$, $1\cdot 10^{-3}$ and $1\cdot 10^{-2}$,
respectively.
 
As in the previous cases if we assume that $\p_\t(p)$ is a Gaussian
we see that the results are:
 
$$\eqalign{
&A=29.2\pm 0.002 \quad{\rm (Gaussian \ assumption)}\cr
&A=30.7\pm 0.1 \quad{\rm (experiment)}\cr
}\Eq(4.23)$$
%
%The data are not very many as the experiment is much harder (in terms of
%cpu time); we report them only because they again seem to suggest that
%distribution is Gaussian, although we know it is not.
\*
\0{\it\S5. Error analysis of the distribution $\p_\t(p)$.}
\numsec=5\numfor=1\*
 
A brief description of the methods that we use to define the errors
follows.
 
For a given $\t$, we build a time sequence of different $p$ values:
$$p_n\equiv p(x_{nt'})=\fra{\s_\t(x_{nt'})}{\langle\s \rangle_+^M}
\kern1.truecm
n=1,....M \Eq(5.1)$$
where $x_{nt'}=S^{nt'}x$, $t'=\t/2+\D$ with $x$ randomly chosen in phase
space with absolutely continuous distribution.and we have chosen $\D=50$
in order to decorrelate contiguous evolution data points. The $\langle\s
\rangle_+^M$ is fixed by normalization, so that the property
$M^{-1}\sum_{n=1}^M p(x_{nt'})=1$ holds, see \equ(2.7).
 
That $\D=50$ is a sufficient delay is warranted by the size of the
entropy autocorrelation decay rate (denoted $\th$ in the tables of \S2).
In general we would expect that $\D$ should be large compared to
$\th^{-1}$. An accurate determination of $\th$ (including a
verification of the validity of an exponential law of decay, which is
not \ap obvious: see the similar problems arising for the one particle
case in [GG]) is a major enterprise and we have only made a few
empirical tests on the order of magnitude of $\th$: and our choice
$\D=50$ was dictated by purely numerical reasons as a reasonable
compromise between small statistics, accuracy and computer availability.
It is justified, in all our experiments, only by the empirically
determined independence of the results (apart from the size of the
errors) in the cases $\D=20$ and $\D=50$. We reported only the $\D=50$
results.
 
We define the discrete probability distribution:
$$\p_\t^D(l;M)={{\sum_{p_n\in I(l)}1}\over{M}}
\kern2.truecm l=-100,..,100\Eq(5.2)$$
where $I(l)=[lr,(l+1)r]$ and $r=1/10$.  The relation between the
discrete and the continuous distribution when $M\rightarrow\infty$ is:
$$\p_\t^D(l;\infty)=\int_{I(l)} dp \p_\t(p)  \Eq(5.3)$$
Assuming that $r$ is small enough we can expand the right hand side of
\equ(5.3):
$$r^{-1}\p_\t^D(l;\infty)=\p_\t(lr+r/2)+{{r^2}\over{24}}{{\dpr^2 \p_\t(p)}
\over{\dpr p^2}}\vert_{p=lr+r/2}+O(r^4) \Eq(5.4)$$
We have checked that this $O(r^2)$ correction is in all our cases
negligible compared to other sources of error: this has been done by
approximating in \equ(5.3) the function $\p_\t(p)$ by a Gaussian
distribution (which, \aps, is a good approximation to $\p_\t(p)$ in the
range of $p$'s that we study):
$$\p_\t(p)\simeq
{{1}\over{\sqrt{2\p}\s}}\exp\left\{-{{(p-1)^2}\over{2\s^2}} \right\}
\Eq(5.5)$$
where $\s\equiv m_2$ is the experimental standard deviation of the
distribution.  By substituting \equ(5.5) into the term with the
second derivative in \equ(5.4) we get:
$$\p_\t(p)=r^{-1}\p_\t^D(l;\infty)\left\{1-{{r^2}\over{24\s^2}}\left[
{{(p-1)^2}\over{\s^2}}-1\right]+O(r^4)\right\} \Eq(5.6)$$
with $p=lr+r/2$.
 
We also have to estimate the error involved in approximating
$\p_\t^D(l;\infty)$ by $\p_\t^D(l;M)$. Let us define
$$P_{\t,l}(m;M)={M\choose m}\p_\t^D(l;\infty)^m \left(1-\p_\t^D(l;\infty)
\right)^{M-m} \Eq(5.7)$$
where $P_{\t,l}(m;M)$ is the probability that in $M$ elements of a
sequence there exists $m$ in the interval defined by $l$. This means
that we regard the various measurements of $m$ as {\it independent}:
because the empirical autocorrelation of the the entropy production
decays on scales smaller than $\D$, the interval between successive
measurements.
 
Then, the average value of $m$ and its mean square displacement are
given by:
$$\langle m\rangle_{\t,l}=M \p_\t^D(l;\infty)$$
$$\langle (m-\langle m\rangle_{\t,l})^2\rangle={{\p_\t^D(l;\infty)
   \left[1-\p_\t^D(l;\infty) \right]}\over{M}} \Eq(5.8)$$
When $M$ is large enough, we shall assume that:
$$\p_\t^D(l;\infty)=\p_\t^D(l;M)\pm 3\left[{{\p_\t^D(l;M)\left(
1-\p_\t^D(l;M)\right)} \over{M}}\right]^{1/2} \Eq(5.9)$$
is an appropriate measurement of the error.
Equations \equ(5.6) and \equ(5.9) are the relations used in
the distribution analysis.
 
The error analysis of the fits for $m_2$, $m_3$, $\k$ is the same as the
one used  in [GG] and it is repeated, with some minor changes to adapt
it to the present sistuations, in the Appendix below.
 
\*
\0{\it\S6. Conclusions. Outlook.}
\numsec=6\numfor=1 \*
 
\0{\it(i) Consistency and CODES}: The above experiments, see Fig.
4,6,9,12 seem in ``good'' agreement with the theoretical predictions. We
have attempted at a very accurate test compatibly with our computer
resources and the need of a natural time cut-off on the duration of the
experiments. The only real limitation was the computer time available;
not so much as available to us but to present day technology. By using
the largest existing computers our results do not seem to be
substantially improvable: the motion being chaotic there is not much
that one really can do without really new ideas.
 
Our attitude is that the theory is general and it should apply to any
system like the ones described in \equ(2.2). So {\it in particular} to
our computer programs that we can refer to as the CODES. For such
dynamical systems our experiments are, by definition exact and the
systems are also by construction ``close'' to \equ(2.2).
 
Therefore on such grounds we could say that we can think that the only
errors in our theory are the statistical errors, \ie our experiment is
as ``perfect'' as one could wish.
 
Nevertheless, not surprisingly, the situation is more subtle. The
reason is mainly that we have been unable to write CODES which ``solve''
\equ(2.2) (in some sense, that is not very relevant for us here) and
which at {\it at the same time verify the time reversibility property}.
 
We think that this is a serious flaw: because the theory, see [GC2],
{\it rests} on time reversal. Strictly speaking, then, we should apply
the first part of the chaotic hypothesis and only assume that the
attractor verifies the Axiom A.
 
However for this we have no theory: the only argument that one could
give, and which we do not find convincing, is that the CODES are
certainly ``close'' approximations to \equ(2.2). And we are used to
think that close systems behave closely. The abundance of
counterexamples has not deterred people to have the feeling that there
is some truth to such belief ({\it natura non facit saltus}).
 
But since it might be simply impossible to write reversible, energy
preserving,  CODES what we have done looked to us to be the only
possibility we had to test the principle.
 
It would be interesting to carry out experiments analogous to the above
described ones for other types of systems for which reversible
algorithms can be written and implemented at least in the conservative
case, as discovered in [LV]: it is unclear, however, if they can be
extended to cover the dissipative cases and, if so, if this can be done
with the accuracy necessary to the test of the chaotic
hypothesis.\annota{4}{\nota The important paper [LV] had escaped our
attention until very recently, too late to take it into account: it
proves that reversible codes do exist for systems not too far from ours
and it is certainly important, and probably possible, to try to adapt
them to test the chaotic hypothesis in a truly reversible CODES.} \*
 
\0{\it(ii) The pairing rule, Axiom A attractors and reversibility:} If
the chaoticity hypothesis, in the form given in \S1 for the reversible
case, is interpreted as meaning that the system behaves as a transitive
Anosov reversible system, with the time reversal operation being the
"global time reversal" $i$, then it implies that the stable and
unstable manifolds have the {\it same dimension}.  So we would expect
to have as many positive and negative Lyapunov exponents.
 
However the existence of the phenomenon in which as $E$ grows some Lyapunov
exponent that is $>0$ at small $E$ becomes $<0$ at larger $E$ was
already pointed out in [DPH].  And tests showed, see Fig.14 and Fig.15
below, that in the case of $10$ particles with semi-periodic boundary
conditions the $19$-th (out of the $38$) Lyapunov exponents is likely
to be $<0$.

\dimen2=25pt \advance\dimen2 by 198pt
\dimen3=276pt \advance\dimen3 by 70pt
\eqfig{\dimen3}{223pt}{
\ins{306pt}{0pt}{$\hbox to20pt{\hfill ${\scriptstyle j}$\hfill}$}
\ins{24pt}{\dimen2}{$\hbox to25pt{\hfill ${\scriptstyle \lambda}$\hfill}$}
\ins{33pt}{3pt}{$\hbox to 12pt{\hfill${\scriptstyle 0}$\hfill}$}
\ins{46pt}{3pt}{$\hbox to 12pt{\hfill${\scriptstyle 1}$\hfill}$}
\ins{59pt}{3pt}{$\hbox to 12pt{\hfill${\scriptstyle 2}$\hfill}$}
\ins{72pt}{3pt}{$\hbox to 12pt{\hfill${\scriptstyle 3}$\hfill}$}
\ins{85pt}{3pt}{$\hbox to 12pt{\hfill${\scriptstyle 4}$\hfill}$}
\ins{97pt}{3pt}{$\hbox to 12pt{\hfill${\scriptstyle 5}$\hfill}$}
\ins{110pt}{3pt}{$\hbox to 12pt{\hfill${\scriptstyle 6}$\hfill}$}
\ins{123pt}{3pt}{$\hbox to 12pt{\hfill${\scriptstyle 7}$\hfill}$}
\ins{136pt}{3pt}{$\hbox to 12pt{\hfill${\scriptstyle 8}$\hfill}$}
\ins{149pt}{3pt}{$\hbox to 12pt{\hfill${\scriptstyle 9}$\hfill}$}
\ins{161pt}{3pt}{$\hbox to 12pt{\hfill${\scriptstyle 10}$\hfill}$}
\ins{174pt}{3pt}{$\hbox to 12pt{\hfill${\scriptstyle 11}$\hfill}$}
\ins{187pt}{3pt}{$\hbox to 12pt{\hfill${\scriptstyle 12}$\hfill}$}
\ins{200pt}{3pt}{$\hbox to 12pt{\hfill${\scriptstyle 13}$\hfill}$}
\ins{213pt}{3pt}{$\hbox to 12pt{\hfill${\scriptstyle 14}$\hfill}$}
\ins{225pt}{3pt}{$\hbox to 12pt{\hfill${\scriptstyle 15}$\hfill}$}
\ins{238pt}{3pt}{$\hbox to 12pt{\hfill${\scriptstyle 16}$\hfill}$}
\ins{251pt}{3pt}{$\hbox to 12pt{\hfill${\scriptstyle 17}$\hfill}$}
\ins{264pt}{3pt}{$\hbox to 12pt{\hfill${\scriptstyle 18}$\hfill}$}
\ins{277pt}{3pt}{$\hbox to 12pt{\hfill${\scriptstyle 19}$\hfill}$}
\ins{289pt}{3pt}{$\hbox to 12pt{\hfill${\scriptstyle 20}$\hfill}$}
\ins{0pt}{33pt}{$\hbox to 26pt{\hfill${\scriptstyle -0.200}$\hfill}$}
\ins{0pt}{53pt}{$\hbox to 26pt{\hfill${\scriptstyle -0.150}$\hfill}$}
\ins{0pt}{73pt}{$\hbox to 26pt{\hfill${\scriptstyle -0.100}$\hfill}$}
\ins{0pt}{93pt}{$\hbox to 26pt{\hfill${\scriptstyle -0.050}$\hfill}$}
\ins{0pt}{114pt}{$\hbox to 26pt{\hfill${\scriptstyle -0.000}$\hfill}$}
\ins{0pt}{134pt}{$\hbox to 26pt{\hfill${\scriptstyle 0.050}$\hfill}$}
\ins{0pt}{154pt}{$\hbox to 26pt{\hfill${\scriptstyle 0.100}$\hfill}$}
\ins{0pt}{174pt}{$\hbox to 26pt{\hfill${\scriptstyle 0.150}$\hfill}$}
\ins{0pt}{194pt}{$\hbox to 26pt{\hfill${\scriptstyle 0.200}$\hfill}$}
\ins{308pt}{183pt}{$\hbox to 26pt{\hfill${\scriptstyle 0.000}$\hfill}$}
\ins{308pt}{197pt}{$\hbox to 26pt{\hfill${\scriptstyle 0.002}$\hfill}$}
\ins{312pt}{169pt}{$\hbox to 26pt{\hfill${\scriptstyle -0.002}$\hfill}$}
\ins{312pt}{156pt}{$\hbox to 26pt{\hfill${\scriptstyle -0.004}$\hfill}$}
\ins{312pt}{142pt}{$\hbox to 26pt{\hfill${\scriptstyle -0.006}$\hfill}$}
}{figliap0}{\eq (pair)}
\vskip 0.5cm
\didascalia{\it Fig.14: The
$38$ Lyapunov exponents for $10$ and $\fra12(pbc)$. The small picture
is an enlargment of the tail of the larger one and it shows more 
clearly the pairing rule and that the $19$--th exponent is slightly 
negative.}

A simple test to see how reliably we have evaluated the Lyapunov
exponents can be, nevertheless, easily made: it consists in testing the {\it
pairing rule}, discovered in [EM], [ECM1], verified to an almost
unthinkable accuracy in experiments by [DPH],  and recently proved for
cases that include precisely our systems, see [DM].  Although we did
not push our study to check the pairing with the remarkable precision
reached in [DPH] we find that the pairing rule is obeyed.  

We stress, however, that the program we use to compute the Lyapunov
exponents is very close (although not identical) to the actual scheme
followed in [DM] to {\it prove} the pairing rule: therefore the test
does not provide a test on the accuracy of the exponents.  The pairing
rule has in fact to be fullf\/illed with high precision even if the
precision reached in measuring the exponents is not comparable (in
other words the errors that one makes on each exponent of a pair
compensate exactly, see [DM], if they are due to the shortness of the
runs).

The existence of one negative exponent in excess
lead us to study the question in more detail. 

In Fig.15 is shown the graph of the Lyapunov exponents $\l_j$, with
$j=13\div19$ in the $\fra12$pbc and $10$ particles for $E=0.1$ to
$E=5.0$ at steps of $0.1$.  The results are still "raw" in the sense
that we have not yet been able to make a satisfactory study of the
errors: the method we used is the ``usual method", [BGGS], with a test
trajectory of $0.5\cdot10^5$ collisions.  The errors are however quite
large and more accurate experiments are necessary to confirm the raw
data.
\1 

\dimen2=25pt \advance\dimen2 by 198pt
\dimen3=276pt \advance\dimen3 by 40pt
\eqfig{\dimen3}{223pt}{
\ins{306pt}{0pt}{$\hbox to20pt{\hfill ${\scriptstyle E}$\hfill}$}
\ins{24pt}{\dimen2}{$\hbox to25pt{\hfill ${\scriptstyle \lambda}$\hfill}$}
\ins{36pt}{3pt}{$\hbox to 23pt{\hfill${\scriptstyle 0.0}$\hfill}$}
\ins{60pt}{3pt}{$\hbox to 23pt{\hfill${\scriptstyle 0.5}$\hfill}$}
\ins{83pt}{3pt}{$\hbox to 23pt{\hfill${\scriptstyle 1.0}$\hfill}$}
\ins{107pt}{3pt}{$\hbox to 23pt{\hfill${\scriptstyle 1.5}$\hfill}$}
\ins{130pt}{3pt}{$\hbox to 23pt{\hfill${\scriptstyle 2.0}$\hfill}$}
\ins{154pt}{3pt}{$\hbox to 23pt{\hfill${\scriptstyle 2.5}$\hfill}$}
\ins{177pt}{3pt}{$\hbox to 23pt{\hfill${\scriptstyle 3.0}$\hfill}$}
\ins{201pt}{3pt}{$\hbox to 23pt{\hfill${\scriptstyle 3.5}$\hfill}$}
\ins{224pt}{3pt}{$\hbox to 23pt{\hfill${\scriptstyle 4.0}$\hfill}$}
\ins{248pt}{3pt}{$\hbox to 23pt{\hfill${\scriptstyle 4.5}$\hfill}$}
\ins{271pt}{3pt}{$\hbox to 23pt{\hfill${\scriptstyle 5.0}$\hfill}$}
\ins{0pt}{19pt}{$\hbox to 26pt{\hfill${\scriptstyle -0.025}$\hfill}$}
\ins{0pt}{42pt}{$\hbox to 26pt{\hfill${\scriptstyle -0.020}$\hfill}$}
\ins{0pt}{65pt}{$\hbox to 26pt{\hfill${\scriptstyle -0.015}$\hfill}$}
\ins{0pt}{88pt}{$\hbox to 26pt{\hfill${\scriptstyle -0.010}$\hfill}$}
\ins{0pt}{111pt}{$\hbox to 26pt{\hfill${\scriptstyle -0.005}$\hfill}$}
\ins{0pt}{133pt}{$\hbox to 26pt{\hfill${\scriptstyle 0.000}$\hfill}$}
\ins{0pt}{156pt}{$\hbox to 26pt{\hfill${\scriptstyle 0.005}$\hfill}$}
\ins{0pt}{179pt}{$\hbox to 26pt{\hfill${\scriptstyle 0.010}$\hfill}$}
\ins{0pt}{202pt}{$\hbox to 26pt{\hfill${\scriptstyle 0.015}$\hfill}$}
}{figl0}{\eq (film)}
\vskip 0.5cm
\didascalia{\it Fig.15: The $13$--th through the $19$-th Lyapunov
exponents in the case $N=10$, $\fra12(pbc)$: row data (no error bars
estimated).  The pairing rule, [EM], [ECM1], [SEM], [DM] is verified.}

The preponderance of the negative Lyapunov exponents over the positive
ones does not seem due, at least in the cases where it appears clearly,
to the non-reversibility of our CODES: hence the existence of more
negative than positive exponents is a very likely to be a real
phenomenon at large field (and fixed $N$).
 
Then one can ask the question: is there any reason to think that the
fluctuation theorem might be valid even if the attractor is ``only'' an
Axiom A attractor?
 
We have not been able to test the fluctuation law at values of $E$ where
there was an apparent excess (we say ``apparent'' because our error
analysis in not satisfactory, yet) of negative Lyapunov exponents, (say
at $E=2.5$ in the case of $10$ particles and $\fra12(pbc)$ where we see
$2$ negative Lyapunov exponents in excess, see Fig.15). The reason is
that the estimated time for the corresponding numerical experiment is
well beyond the present day computer capacities. Large fluctuations may
be very difficult to observe at very large fields.
 
We do not think that the fluctuation law can hold by chance: but we
have seen it well verified in a situation where there seem to be one
negative Lyapunov exponents in excess over the positive ones, see Fig.12 
(and Fig.15 with $E=1$).  However from the analysis in [GC1],
[GC2] its validity appears very tightly related to the existence of a
time reversal symmetry {\it leaving the closure of the attractor
invariant}, \ie a map $i^*$ defined on the closure of the attractor and
such that $S i^*=i^* S^{-1}$.
 
Therefore we conclude  that a not unreasonable scenario would be that
when there are pairs of Lyapunov exponents that consist of two {\it
negative} exponents then the attractor and its accumulations points can
be simply regarded as a {\it smooth lower dimensional
surface}.\annota{5}{\nota This does not preclude the possibility that the
attractor has a fractal dimension (smoothness of the closure of an
attractor has nothing to do with its fractal dimensionality, see
[ER],[GC1],[G1]).}  The motion on this lower dimensional surface (whose
dimension is smaller than that of phase space by an amount equal
to the number of paired negative exponents) will still have an
attractor (with dimension {\it lower} than the dimension of the surface
itself, as suggested by the Kaplan--Yorke formula, [ER]).  And on such
manifold the motion will still be {\it reversible} in the sense that
there will be a map $i^*$ of the attractor into itself ({\it certainly
different from} the global time reversal map $i$) which {\it inverts}
the time on the attractor and that can be naturally called a {\it local
time reversal}, see [BG].
 
Then manifestly one would be back with an Anosov system (on a lower
dimensional manifold) and a version of the fluctuation theorem would
still hold. Furthermore one could say that this is only a different
interpretation of the chaotic principle of [GC1], [GC2] (which in such
case does not even require to be reformulated to apply).
 
If this picture is correct we can write the phase space contraction rate
(see \equ(2.4),\equ(2.5)) $\s(x)=\s_0(x)+\s_\perp(x)$ where $\s_0(x)$ is
the contraction rate on the surface on which the attractor lies and
$\s_\perp(x)$ is the contraction rate of the part of the stable manifold
of the attractor which is not on the attractor itself (the angle between 
the part of the stable manifold sticking out of the attractor and the 
attractor itself is disregarded here as we think that it is bounded away 
from $0$ and $\p$ since the attractor is compact).
 
Of course the local time reversal will change the sign {\it only of}
$\s_0(x)$ and the fluctuation theorem should apply to the fluctuations
of $\s_0$. But $\s_0(x)$ is {\it not} directly accessible to
measurement: nevertheless we can still study its fluctuations via the
following {\it heuristic} analysis. From the proof in [DM] of the
pairing rule one sees that the jacobian matrix $J$ of the map $S$ is
such that $\sqrt{J^*J}$ has $D$ pairs of eigenvalues and the logarithms
of each pair add  up to $\ig_0^{t(x)} \a(Q_tx)dt$ (see also
\equ(2.4),\equ(2.5)).
 
The simplest interpretation of this, in view of the above proposed
picture of the attractor, is that the pairs with elements of opposite
signs describe expansion {\it on the manifold} on which the
attractor lies. While the $M\le D$ pairs consisting of two negative
eigenvalues describe the contraction of phase space in the directions
{\it transversal to the manifold} on which the attractor lies. Then we
would have $t_0\s_0(x)=(D-M)\ig_0^{t(x)}\a(Q_tx)dt$ and we should have a
fluctuation law for the quantity $p(x)$ associated with $\s_0(x)$
defined by \equ(2.6) and \equ(2.7) with $\s_0$ replacing $\s$, \ie
(accepting the above heuristic argument):
$$\s_{0\t}(x)=\fra{(D-M)}D\media{\s}_+ p(x)\Eq(6.2)$$
\ie a law identical to \equ(2.10) {\it up to a correcting factor}
$1-\fra{M}D$:
$$\log\fra{\p_\t(p)}{\p_\t(-p)}=\t\, t_0\media{\s}_+\,
(1-\fra{M}D)\,p\Eq(6.3)$$
The graphs of Fig.12 are relative to an experiment in which we see that
there may be one negative exponents in excess over the positive ones
(as said above, see also Fig.15): it is very small (see Fig.14), 
and it carries an error bar that we estimate to be so large to allow
for positive values as well. The graphs, however, show
that the agreement with the  experiment of \equ(6.3) is within the
errors: had there been no negative exponents we would have expected a
slope $1$. If there is one negative exponent in excess we expect a slope
$1-\fra1{19}$ which is within the error bars in Fig.12
(had we drawn in Fig.12 the best fit line rather than the line with 
slope $1$ the agreement with the slope $1-\fra1{19}$ would have been 
even better).  An excess of $2$ exponents would yield a slope of
$1-\fra2{19}$ which is {\it out} of the error bars. 

Note that since the exponent smallest in modulus is
so small we must expect that it yields a clear effect only after
extremely long times have elapsed (\ie for values of $\t>4.\,10^3$:
totally out of computability). 
 
The difficulty on the above scenario is that there is no \ap reason to
think that the attractors should have the above structure: \ie fractal
sets lying on smooth surfaces on phase space {\it on which the motion is
reversible}. However such a picture is very suggestive and it might be
applicable to more general situations in which the reversibility holds
only {\it on the closure of the attractor} and not in the whole space
(like "strongly dissipative systems"). An attempt at discussing this
point can be found in [BG].
 
%In the comment {\it(v)} below we present one more experimental result which
%seems to suggest (indirectly) that \equ(6.3) may hold in a situation in
%which the field is so large that we are reasonably sure that there is
%one negative Lyapunov exponents in excess over the positive ones
%($N=2,E=5$ and $\fra12pbc$).
\*
 
\0{\it(iii) Smoothness of the closure of the attractor:} The smoothness
of the surface on which the attractor lies, so that the attractor can be
regarded, in itself, as an Anosov system, is very likely not necessary,\
from a  mathematical viewpoint. The more general assumption that the
attractor verifies the Axiom A (and is transitive) would be sufficient
if accompanied by the ({\it very strong}) assumption that the attractor
is mapped into itself by a symmetry $i^*$ such that on the attractor
$i^* S=S^{-1} i^*$ (discussed in (ii) above). We stress again  that
$i^*$ {\it cannot be the same} as the global time reversal $i$ because the
latter will map the attractor for the forward motion into the one for
the backward motion: however through simple examples of reversible
motions with attractors closures  smaller than the whole phase space
(hence verifying Axiom A but not the Anosov property) one can see that
the existence of $i$ often (always?) induces the existence of a map
$i^*$ on the attractor, se [BG].
 
We insist in talking about smoothness for the following two reasons:
 
(1) because we think that the theory applies to general many particle
systems and we cannot see the relevance of a possible fractional
dimension over $N=10^{19}$ total dimensions: in other words we can
proceed as if the dimension was integer (in [GC2] the fractality is
called ``an unfortunate accident'' that may happen in the problems that
we study), hence as if the system is Anosov, {\it provided we accept that the
attractor has a time reversal symmetry} (\ie there is a map $i^*$ of the 
attractor into itself that anticommutes with the time evolution, 
$S_ti^*=i^*S_{-t}$).
 
(2) because, as it partially appears from Fig.15, the Lyapunov
exponents seem to evolve, as a function of $E$, along the following
pattern.  At smaller $E$ they are all non zero (as shown by Fig.15):
the dimension of the closure of the attractor in $4N-2$ (\ie that of
the phase space $\CC$).  Then as $E$ grows one of them crosses
continuously (and with non zero $E$ derivative, \ie transversally) the
value $0$ at some $E_1>0$ becoming negative for larger values of the
field; the closure of the attractor has now dimension $4N-4$.  At
$E_2>E_1$ a second Lyapunov exponent crosses $0$ (transversally) and
the closure of the attractor has now dimension $4N-6$, \etc.  This
suggests that, as $E$ varies, the attractor is characterized by more
and more ``constants of motion'', \ie by the vanishing of more and more
observables.  Every time one more ``constant of motion'' is born we see
that the attractor loses two dimensions.  Nothing suggests to us that
it becomes non smooth, or appreciably so.
 
One should also bear in mind that the above analysis is, necessarily,
carried over in systems with few degrees of freedom. It might well be
that the picture can considerably simplify at large $N$. See below for
some thoughts on that point.  \*
 
\0{\it(iv) Fluctuation theorem as a reversibility test:} Since the very
derivation, [GC1],[GC2],[G2], of the fluctuation theorem is so
intimately related to reversibility one could say the if the predictions
of the fluctuation theorem are verified, perhaps with a slope $<1$ as
suggested by \equ(6.3), then this is a sign that the
dynamics on the attractor is reversible in the sense that it is mapped
into itself by a map $i^*$ such that $i^*S=S^{-1}i^*$ (hence $i\ne i^*$
unless the system is Anosov and the attractor is the whole space). It is
clear that the above considerations give rise to several tests that can
be experimentally performed. 

Note that not only the linearity in $p$ is a strong statement, but also 
the predicted value $<1$ is somewhat surprising: naively one could be 
tempted to think that the contraction rate transversal to the attractor 
is {\it uniform} over the attractor: this would lead to a slope $>1$
(and the larger the stronger is the attraction from the 
attractor).
\*
 
\0{\it(v) Gaussian? Central limit theorem?} One may say, as a
first reaction to the analysis, ``of course we expected a Gaussian
distribution of the fluctuations because the dissipation $\t\s_\t$ is a
sum of many terms that are statistically independent'' if the motion is
chaotic. Hence ``everybody (reasonable)'' would expect a Gaussian
distribution for the fluctuations. This would mean that $(p-1)$ has a
dispersion of order $C\t^{-\fra12}$ for some (non trivial) constant $C$
related to the entropy autocorrelation function
$\media{\s(S^nx)\s(x)}$:
$$C^2=\fra1{\media{\s}_+^2}\sum_{n=-\io}^{\io}
\big(\media{\s(S^n\cdot)\s(\cdot)}_+-\media{\s}_+^2\big)\Eq(6.4)$$
And it would imply that:
$$\log\fra{\p_\t(p)}{\p_\t(-p)}=\t \fra2{C^2}\,p\Eq(6.5)$$
However this {\it would not} imply that $\fra2{C^2}= t_0\media{\s}_+$:
which would mean that, {\it in general}, it is {\it also}:
$$\media{\s_+}= \fra{t_0}2\sum_{n=-\io}^{\io}
\big(\media{\s(S^n\cdot)\s(\cdot)}_+-\media{\s}_+^2\big)\Eq(6.6)$$
The latter would be a {\it strange} prediction in the context of the central
limit theorems. Furthermore the central limit theorem is expected to
hold only for fluctuations of $p-1$ the order of $\t^{-1/2}$ while
\equ(2.10) compares the probabilities of $p$ to those of $-p$ which, for
$p\simeq1$, describe deviations of order $\t$. In other words there is
{\it no relation} between the usual central limit theorem
for $p$ and the fluctuation theorem (when they are expected to hold,
\ie for $\t\to\io$).
 
Therefore it came as a {\it surprise} to us that the results were {\it
instead} consistent, on the whole range of observability, with a
Gaussian distribution. We would have thought that
$\log\p_\t(p)=-\t\z(p)$ (for $\t$ large enough) with only the odd part
of $\z(p)$ {\it strictly proportional to $p$}, while the even part had
no reason to be quadratic. And the example of \S3 shows how natural it
would be that is not quadratic. The constant $C$ would have been mainly
determined by the curvature of $\z(p)$ at its maximum $p=1$, totally
unrelated (we thought) to $\fra12t_0\media{\s}_+$.
 
Naturally we expected \equ(6.6) to hold for very small external field
$E$: in the case $N=1$ it was indeed proved by [CELS] to hold for values
of $E$ that, in our units, are {\it extremely small} compared to $1$:
and the equation \equ(6.6) is known for small field (``linear regime'')
as the "fluctuation dissipation" relation (or ``Green--Kubo formula'').
This relation has been shown to be closely related (and essentially a
consequence) of the chaotic hypothesis in reversible systems for $E$
close to $0$, (see [G5] eq. (5.9), taking into account the special form
of $\s$, its linearity in $E$ and the fact that the energy is $1$) still
with corrections of $O(E^3)$.
 
Note that as $E\to0$ the two sides of \equ(6.6) have both size
of order $O(E^2)$ and the fluctuation dissipation theorem is the
relation obtained by dividing the two sides by $E^2$ and letting
$E\to0$.
 
There should be a deeper reason for the relation between the small
deviations constant $C$ (relevant in the fluctuations of scale $\sqrt
\t$) and the large scale fluctuations (of size $O(\t)$) which are
related to $\media{\s}_+$, see \equ(2.10): this seems to be explained in 
[G6].
 
In the ``old'' literature one can find statements that today sound
somewhat mysterious like:
\*
\0{\it ``It is empirically known that for macroscopic values of $\V\a$,
\ie for values of the $\a_i$ much larger than their root mean square
values at equilibrium, the averages of these quantities frequently obey
linear differential equations.''} \*
 
\0which is then used to establish that relations between properties that
hold for small fluctuations hold also for large ones, at least in the
average. In [DGM], p. 100, the above statement is the beginning of a
classical derivation of the Onsager reciprocity relations. One may
speculate, since the chaotic hypothesis is related to the Onsager
relations as well, [G5], that our ``strange'' relation between the small
and large deviations might be related to the fact that a similar
property can be taken as the basis of a derivation of Onsager's theory
of reciprocity. Small and large fluctuations seem to have some common
properties that one may not expect \ap or, at least, that one may
consider worth of being challenged and tested, see [DGM], p. 102.
This point of view leads to the paper [G6] where it is developed and its 
consequences are drawn.
 
The connection between large and small fluctuations is, if at all
present, likely to be a peculiarity of models typical of statistical
mechanics (short range interacting many particle systems): it may not
hold in other cases in which the chaotic hypothesis can be applied
(sometimes leading, nevertheless, to reciprocity relations of Onsager
type) like in fluidodynamics models, [G4].
 
It is worth stressing again that we {\it know} that the distribution in $p$
cannot be Gaussian for all $p$'s: because there is a maximum
($\t$--independent) value that $p$ can take, just from the finiteness of
phase space. This value, called $p^*$ in [G2], can be easily measured in
our experiments (see \equ(2.12)) but it is very far away from the region
where we have enough statistics to make meaningful measurements, as it
appears from the graphs reported above.
\*
 
\0{\it(vi) Time scales for large $N$:} A somewhat more speculative
scenario can be drawn for large $N$. We mention it because we hope to
receive some help in a program directed to test it. It seems reasonable
to us that {\it in the thermodynamic limit} the Lyapunov exponents of
the system will fall into two categories each consisting of a pair of
exponents. The sum of the values of each pair is {\it constant} and
equal to half the average entropy creation rate (this is the {\it
pairing rule}, see above).
 
A large number of pairs ($O(N)$) will consist of one vanishing exponent
and its negative companion. {\it The remaining positive Lyapunov
exponents will all be identical: marking the time scale of local approach
to equilibrium}, hence also the other negative exponents will be
identical (by the pairing rule). By {\it identical} we mean here that
the ratio between the largest and the smallest positive Lyapunov
exponents is bounded uniformly in $N$ (away from $0$ and $\io$).
 
This is in perfect agreement with fig. 5 of [DPH] describing a {\it very
high density system}: it is {\it not} in agreement with
the other results of [DPH].
 
Nevertheless, as argued in also in [G3], it is possible that the low density
results are flawed in this respect as one would need far too large
systems to obtain Lyapunov exponents for a distribution close to the
thermodynamic limit distribution $\m$. Only in the high density case a
small sample of gas exhibits the features of a large sample.
 
The above picture merges with the ideas in [G4] relating the $0$
Lyapunov exponents to the macroscopic modes described by macroscopic
equations, while the nonzero exponents describe the approach to local
equilibrium.  It also matches with Fig.14 where a rather sharp drop of
the Lyapunov exponents towards $0$ appears betwen the $10$--th and the
$11$--th exponent.

Since it is always assumed that there is {\it only one
microscopic time scale} for the local approach to equilibrium it is
perhaps a natural conjecture that the Lyapunov exponents should have the
above structure.\*
 
\0({\it vii}) We see that the above experiments raise perhaps more problems
than expected: but the chaotic hypothesis emerges as not inconsistent
with the data.
 
\*
{\it\S Appendix. Fits and Errors.}
\numsec=1\numfor=1
\*
This appendix is adapted to the present experiments from the
corresponding appendix in [GG].  We get data sets from our computer
experiments, say $\V y(\V x)=
\{y(x_i)\}_{i=1,N}$, where $\V x=\{x_i\}_{i=1,N}$ is in our case an
independent variable, say for instance a set of $N$ collision numbers or
time instants.  To the latter experimental data set of points, we want
to {\it fit} a given {\it guessed} function, say $f(x;\V\a)$, where
$\V\a=\{\a_n\}_{n=1,p}$ is a set of arbitrary parameters.  Here by {fit}
we mean to find a set of parameters $\V\a^{*}$ which {\it optimizes}
some reasonable functional relation between the experimental data and
the fitting function.
 
In our case we use the {\it least squares functional}, \ie :
$$V(\V y(x),\V\a)=\sum_{i=1}^{N}\,\left [ y_i - f(x_i;\V\a) \right ]^2
 \Eqa(A1.1)$$
The set of parameters $\V\a^{*}(\V y)$ is here obtained by asking that
they should be the minima of the $V$ function: $\partial_{\V\a^{*}}V(\V
y,\V\a^{*}) = 0$ . We also define the {\it goodness} of our fit, $G$, by
the average $y$-distance of our data to the function $f(x;\V\a^{*})$:
$G(\V y(\V x))=\big(V(\V y(\V x),\V\a^{*})/N\big)^{1/2}$. This parameter
is only meaningful when it is compared with the one from another fit.
Given many fits, the one with smallest $G$ value will be called {\it
best fit} (among the considered fits).
 
The data have, in general, non-negligible errors, say $\V\e=
\{\e_i\}_{i=1,N}$, due to the finite number of samples used in the
averaging (see comments in \S2). Such errors induce errors on the
parameter values. Therefore, a measure of the error amplitude in
$\V\a^{*}(\V y)$ is given by:
$$\Delta_{\V\e}\V\a^{*}(\V y)=\left [ \V\a^{*}(\V y+\V\e)
-\V\a^{*}(\V y-\V\e)\right ] /2  \Eqa(A1.2)$$
In the particular case in which the magnitude of the data error is much
smaller than the measured value, $|\e_i/y (x_i)|<<1$, we may expand the
latter equation around $\V\e=\V 0$:
$$\Delta_{\V\e}\a_n(\V y)=\sum_{i=1}^{N}c_{i}^{(n)}(\V y) \e_i \Eqa(A1.3)$$
The coefficients $c_{i}^{(n)}$ are found by expanding $V(\V y(\V
x)+\V\e,\V\a)$ around $\V\a^{*}(\V y)$ and $\V\e=\V 0$ and they are
given by:
$$c_{i}^{(n)}=-\sum_{m=1}^{p} D_{mn}^{-1} \partial_{y(x_i)\a_{m}^{*}}^{2}
V(\V y,\V\a^{*}) \Eqa(A1.4)$$
where $D_{mn}=\partial_{\a_{m}^{*}\a_{n}^{*}}^{2}V(\V y,\V\a^{*})$.
 
In particular for the linear fit, $f(x,\V\a)=\a_1+\a_2\,x$, the
coefficients $c_{i}^{(1),(2)}$ are given by:
$$c_{i}^{(1)}={2\over{N\Delta x}}(\overline{x^2}-\overline{x}x_i),
\qquad c_{i}^{(2)}={2\over{N\Delta x}}(x_i-\overline{x}) \Eqa(A1.5)$$
where $\Delta x = \overline{(x-\overline{x})^2}$ and
$\overline{x^n}=N^{-1}\sum_{i=1}^{N}x_{i}^{n}$.
 
The errors are random variables and we have to average them over their
distribution. In all cases considered it seemed reasonable to consider
the errors $\e_i$ as independent variables. Therefore we {\it
empirically} estimate an upper bound for their correlation values:
$$|\media{\e_i\e_j}|
\le \sqrt{\media{\e_{i}^{2}}\media{\e_{j}^{2}}}\Eqa(A1.6)$$
The parameter errors in our analysis are defined by the equations:
$$\Delta\a_n(\V y)^2=\sum_{i=1}^{N}\sum_{j=1}^{N}c_{i}^{(n)}
c_{j}^{(n)}\delta_{i}\delta_{j}
\ge \media{\Delta_{\V\e}\a_{n}^{2}} \Eqa(A1.7)$$
where $\delta_{i}^{2}=\media{\e_{i}^{2}}$ and their use and meaning
is described entirely by the above comments.
 
\* \0{\it Acknowledgments}: We are indebted to E.G.D. Cohen for his
interest and his comments and suggestions: his constant support and
advice have been very important to us. We thank J. Lebowitz and
D. Ruelle for many discussions during which they provided inspiration
and ideas. This work is part of the research program of the European
Network on: ``Stability and Universality in Classical Mechanics", \#
ERBCHRXCT940460, and it has also been partly supported by Rutgers
University, CNR-GNFM, IHES and MPI.  \*
 
\0{\it References} \*
 
\item{[AA]} Arnold, V., Avez, A.: {\it Ergodic problems of classical
mechanics}, Benjamin, 1966.
 
\item{[Bo]} Bowen, R.: {\it Equilibrium states and the ergodic theory of
Anosov diffeomorphisms}, Lecture notes in mathematics, vol. {\bf 470},
Springer Verlag, 1975.
 
\item{[BG]} Bonetto, F., Gallavotti, G.: {\it Reversibility,
coarse graining and the chaoticity principle}, in preparation.
 
\item{[BEC]} Baranyai, A., Evans, D.T., Cohen, E.G.D.: {\it
Field dependent conductivity and diffusion in a two dimensional Lorentz
gas}, Journal of Statistical Physics, {\bf70}, 1085-- 1098, 1993.
 
\item{[BGGS]} Benettin, G., Galgani, L., Giorgilli, A., Strelcyn,
J.M.: {\it Lyapunov Characteristic Exponents for Smooth Dynamical
Systems and for Hamiltonian Systems; a Method for Computing all of
Them. Part 1: Theory}, Meccanica, {\bf 15}, 9--20, 1980; and {\it
Lyapunov Characteristic Exponents for Smooth Dynamical Systems and for
Hamiltonian Systems; a Method for Computing all of Them. Part 2:
Numerical Applications}.  Meccanica {\bf 15}, 21--30 (1980).

\item{[CELS]} Chernov, N.I., Eyink, G.L., Lebowitz, J.L., Sinai, Y.G.:
{\it Steady state electric conductivity in the periodic Lorentz gas},
Communications in Mathematical Physics, {\bf 154}, 569--601, 1993.

\item{[DM]} Dettman, C.P., Morriss, G.P.: {\it Proof of conjugate pairing
for an isokinetic thermostat}, N.S. Wales U., Sydney 2052, preprint, 
1995.
 
\item{[DPH]} Dellago, C., Posch, H., Hoover, W.: {\it Lyapunov instability in
system of hard disks in equilibrium and non-equilibrium steady states},
preprint, Institut f\"ur Experimentalphysik, Universit\"at Wien,
Austria, August 1995.
 
\item{[DGM]} de Groot, S., Mazur, P.: {\it Non equilibrium thermodynamics},
Dover, 1984.

\item{[EM]} Evans, D.J., Morriss, G.P.: {\it Statistical 
mechanics of nonequilibrium liquids}, Academic Press, 1990.
 
\item{[ECM1]} Evans, D.J.,Cohen, E.G.D., Morriss, G.P.: {\it Viscosity of a
simple fluid from its maximal Lyapunov exponents}, Physical Review, {\bf
42A}, 5990--\-5997, 1990.
 
\item{[ECM2]} Evans, D.J.,Cohen, E.G.D., Morriss, G.P.: {\it Probability
of second law violations in shearing steady flows}, Physical Review
Letters, {\bf 71}, 2401--2404, 1993.
 
\item{[ER]} J. Eckmann, D. Ruelle: {\it Ergodic theory of strange attractor},
Reviews of Modern Physics, {\bf 57}, 617--656, 1993  

\item{[G1]} Gallavotti, G.: {\it Topics in chaotic dynamics}, Lectures
at the Granada school 1994, Ed. P. Garrido, J. Marro,
in Lecture Notes in Physics, Springer Verlag, {\bf 448}, p. 271--311,
1995.
 
\item{[G2]} Gallavotti, G.: {\it Reversible Anosov diffeomorphisms and
large deviations.}, Mathematical Physics Electronic Journal,
{\bf 1}, (1), 1995.
 
\item{[G3]} Gallavotti, G.: {\it Coarse graining and chaos},
in preparation.
 
\item{[G4]} Gallavotti, G.: {\it Chaotic principle: some applications to
developed turbulence}, in {\it mp$\_$arc @math. utexas. edu}, \#95-232,
1995.
 
\item{[G5]} Gallavotti, G.: {\it Chaotic hypothesis: Onsager reciprocity
and fluctuation--dissipation theorem}, in print on J. Statistical 
Physics.

\item{[G6]} Gallavotti, G.: {\it Extension of Onsager's reciprocity to 
large fields and the chao\-tic hypothesis},
mp$\_$ arc 96-109; or chao-dyn 9603003.
 
\item{[GC1]} Gallavotti, G., Cohen, E.G.D.: {\it Dynamical ensembles in
nonequilibrium statistical mechanics}, Physical Review Letters,
{\bf74}, 2694--2697, 1995.
 
\item{[GC2]} Gallavotti, G., Cohen, E.G.D.: {\it Dynamical ensembles in
stationary states}, Journal of Statistical Physics, {\bf 80},
931--970, 1995.
 
\item{[GG]} Garrido, P., Gallavotti, G.: {\it Billiards correlation
functions}, Journal of Statistical Physics, {\bf 76}, 549--586, 1994.
 
\item{[HHP]} Holian, B.L., Hoover, W.G., Posch. H.A.: {\it Resolution of
Loschmidt's paradox: the origin of irreversible behavior in reversible
atomistic dynamics}, Physical Review Letters, {\bf 59}, 10--13, 1987.
 
\item{[L]} Lorenz, E.: {\it A deterministic non periodic flow},
Journal of Atmospheric Science, {\bf20}, 130-- 141, 1963.
 
\item{[LA]} Levi-Civita, T., Amaldi, U.: {\it Lezioni di Meccanica
Razionale}, Zanichelli, Bologna, 1927 (re\-prin\-ted 1974), volumes
$I,II_1,II_2$.
 
\item{[LV]} Levesque, D., Verlet, L.: {\it Molecular dynamics and time
reversibility}, Journal of Statistical Physics, {\bf 72},
519--537, 1993.
 
\item{[M]} Morriss, G.: {\it },
Physics Letters, {\bf 134A}, 307, 1989.
 
\item{[PH]} Posch, H.A., Hoover, W.G.: {\it Non equilibrium molecular
dynamics of a classical fluid}, in "Molecular Liquids: New Perspectives
in Physics and Chemistry", ed. J. Teixeira-Dias, Kluwer Academic
Publishers, p. 527--547, 1992.
 
\item{[R1]} Ruelle, D.: {\it Chaotic motions and strange attractors},
Lezioni Lincee, notes by S.  Isola, Accademia Nazionale dei Lincei,
Cambridge University Press, 1989; see also: Ruelle, D.: {\it Measures
describing a turbulent flow}, Annals of the New York Academy of
Sciences, {\bf 357}, 1--9, 1980.  For more technical expositions see
Ruelle, D.: {\it Ergodic theory of differentiable dynamical systems},
Publications Math\'emathiques de l' IHES, {\bf 50}, 275--306, 1980.
 
\item{[R2]} Ruelle, D.: {\it A measure associated with Axiom A
attractors}, American Journal of Mathematics, {\bf98}, 619--654, 1976.
 
\item{[R3]} Ruelle, D.: {\it Positivity of entropy production in non
equilibrium statistical mechanics}, IHES preprint, 1995.
 
\item{[S]} Sinai, Y.G.: {\it Gibbs measures in ergodic theory}, Russian
Mathematical Surveys, {\bf 27}, 21--69, 1972.  Also: {\it Lectures in
ergodic theory}, Prin\-ce\-ton U.  Press, Princeton, 1977.

\item{[Sm]} Smale, S.: {\it Differentiable dynamical systems}, Bullettin
of the American Mathematical Society, {\bf 73 }, 747--818, 1967.
 
\item{[SEM]} Sarman, S., Evans, D.J., Morriss, G.P.: {\it Conjugate
pairing rule and thermal transport coefficients}, Physical Review,
{\bf42A}, 2233--2242, 1992.
 
\item{[UF]} Uhlenbeck, G.E., Ford, G.W.: {\it Lectures in Statistical
Mechanics}, American Mathematical society, Providence, R.I.,
pp. 5,16,30, 1963.
\*

\0{\it Internet access:
All the Authors' quoted preprints can be found and freely downloaded
(latest postscript version including corrections of misprints and
errors) at:

\centerline{\tt http://chimera.roma1.infn.it}

\0in the Mathematical Physics Preprints page.}
\ciao
\1

\centerline{Figure Captions}

Fig.1: General billiard structure with scatterers of radius
$R_1$ and $R_2$ in a periodic box with side length $L$, (case
$k\times\ell=1\times1$).

Fig.2: The histograms for $\p_\t(p)$ at various values of
the observation time $\t=20,40,60,100$. Each vertical bar is the error
bar centered around the measured point. The dots on the axis mark the
extremes of the interval where the observed data differ form $0$ within
the statistical error.

Fig.3: The decay of the $2^d$ and $3^d$ moments, as
functions of $\fra1\t$ or $\fra1{\t^2}$, and the kurtosis.

Fig.4: The fluctuation theorem test for
$\t=20,40,60,100$. The dashed straight line is the theoretical
prediction of the fluctuation theorem. The arrows mark the point at
distance $\sqrt{\media{(p-1)^2}}$ from $1$. The error bars are inherited
from the histogram data in Fig.2.

Fig.5: The distribution $\p_\t(p)$ for
$\t=20,40,60,100$.

Fig.6: The decay of the $2^d$ and $3^d$ moments, as
functions of $\fra1\t$ or $\fra1{\t^2}$ and of the kurtosis as a
function of $\fra1{\t}$.

Fig.7: The graphs for the fluctuation theorem test for
$\t=20,40,60,100$. The dashed straight line is the theoretical
prediction of the fluctuation theorem. The arrows mark the point at
distance $\sqrt{\media{(p-1)^2}}$ from $1$. The error bars are inherited
from the histogram data in Fig.5.

Fig.8: The histograms for $\p_\t(p)$ for
$\t=20,100$.

Fig.9: The graphs for the fluctuation theorem test, case of
$\fra12(pbc),\,N=2$; the dashed line is the fluctuation theorem
prediction, $\t=20,100$. The arrows mark the point at
distance $\sqrt{\media{(p-1)^2}}$ from $1$.

Fig.10: The second moment, the third moment and the
kurtosis.

Fig.11: Histograms for $\p_\t(p)$ at $\t=20$
and $\t=100$.

Fig.12: The linear fluctuation test, $\t=20,40,80,100$. The
dashed line is the fluctuation theorem prediction for $\t=+\io$. The
arrows mark the point at distance $\sqrt{\media{(p-1)^2}}$ from $1$.

Fig.13: The decay of the $2^d$ and $3^d$ moments and of
the kurtosis.

Fig.14: The
$38$ Lyapunov exponents for $10$ and $\fra12(pbc)$. The small picture
is an enlargment of the tail of the larger one and it shows more 
clearly the pairing rule and that the $19$--th exponent is slightly 
negative.

Fig.15: The $13$--th through the $19$-th Lyapunov
exponents in the case $N=10$, $\fra12(pbc)$: row data (no error bars
estimated).  The pairing rule, [10], [11], [33], [7] is verified.

\ciao